\documentclass[12pt]{article}
\usepackage{graphicx,bm,amsmath,amssymb,latexsym,epsfig,euscript,multirow,color,url,psfrag}
\usepackage{hyperref}
\allowdisplaybreaks

\newcommand{\rd}{\mathrm{d}}
\newcommand{\be}{\begin{equation}}
\newcommand{\ee}{\end{equation}}
\newcommand{\bea}{\begin{eqnarray}}
\newcommand{\eea}{\end{eqnarray}}
\newcommand{\nn}{\nonumber}
\renewcommand{\l}{\left}
\renewcommand{\r}{\right}
\renewcommand{\bar}{\overline}

\newcommand{\vv}{\mathbf}

\newcommand{\cM}{\mathcal{M}}

\newcommand{\go}{{\tilde g}}
\newcommand{\sq}{{\tilde q}}
\newcommand{\st}{{\tilde t}}
\newcommand{\q}{\quad}
\newcommand{\qq}{\qquad}
\newcommand{\qqq}{\qquad\quad}
\newcommand{\qqqq}{\qquad\qquad}
\newcommand{\comment}[1]{}

\input prepictex
\input pictex
\input postpictex
\newdimen\tdim
\tdim=\unitlength
\def\stpltsmbl{\setplotsymbol ({\small .})}

\def\barrow{\arrow <8\tdim> [.3,.6]}

\begin{document}
\begin{titlepage}

\vspace{0.2cm}
\begin{center}
\Large\bf
Annihilation decays of bound states at the LHC
\end{center}

\vspace{0.2cm}
\begin{center}
{\sc Yevgeny Kats and Matthew D. Schwartz}\\
\vspace{0.4cm}
{Department of Physics\\
Harvard University\\
Cambridge, MA 02138, U.S.A.}
\end{center}

\begin{abstract}
\vspace{0.2cm}
\noindent
At the Large Hadron Collider, heavy particles may be produced in pairs close to their kinematic threshold. If these particles have strong enough attractive interactions they may form bound states. Consequently, the bound states may decay through annihilation back into the standard model. Such annihilation decays have the potential to provide much information about the bound particles, such as their mass, spin, or charges, in a manner completely complementary to standard single particle cascade decays. Many of the signatures, such as dijet resonances, will be challenging to find, but may be extremely helpful in unraveling the nature of the new physics. In the standard model, the only novel annihilation decays would be for toponium; these will be hard to see because of the relatively large width of the top quark itself. In models with supersymmetry, marginally visible annihilation decays may occur for example, from bound states of gluinos to dijets or tops. If new particles are bound through forces stronger than QCD, annihilation decays may even be the discovery mode for new physics. This paper presents various theoretical results about bound states and then addresses the practical question of whether any of their annihilation decays can be seen at the LHC.
\end{abstract}
\vfil

\end{titlepage}

\section{Introduction}
It will undoubtedly be much easier to conclude that the LHC has found new physics
than to determine what that new physics is. Many of the most well-motivated models have
signatures such as events with a large number of jets and missing energy. While there is little background from the standard model
for these signatures, the difficultly of sorting out which jet corresponds to which parton, working out the underlying topologies,
reconstructing the intermediate masses, and isolating the relevant couplings may be an insurmountable challenge. Special cases,
for example, with multiple leptons, will make things easier, but determining couplings and masses for strongly interacting particles
may require the precision of a next-generation machine. With no such machine on the horizon, it is therefore imperative to consider other
ways in which parameters of the underlying Lagrangian may be extracted from data at the LHC. In this paper, we consider one
often neglected possibility: massive, strongly interacting particles that are produced in pairs can form bound states which then annihilate back into the standard model.
Such annihilation decays are subject to an entirely different set of systematics than standard cascade decays, and therefore,
if they can be seen, provide the possibility of generating entirely complementary information about the underlying physics.

Bound states exist when the potential between the particles is
attractive, with a binding energy that is larger than the particle's
intrinsic width. If the states are too strongly bound, such as through new
beyond-the-standard model (BSM) forces, we may never see the constituent particles on their own, that is to say, the bound states will look just like resonances. On the other extreme, if the binding energy is much
smaller than the width, the equivalent of bound state production and annihilation is simply virtual pair production, with little evidence of resonance behavior over the continuum. We are interested in the intermediate regime, where the particles sometimes bind and annihilate, and other times have single particle decays, since this is when the two complementary types of signatures can be available. Conveniently, this happens fairly generically if the particles are bound through the strong force $SU(3)$ of the standard model and the single particle decays only go through off-shell intermediate states. In particular, it happens for many standard BSM scenarios, where the bound states will typically amount to a few percent of the cross section of the pair production processes.

If the products of annihilation decays
can be detected and distinguished from the background, they
will provide a wealth of information. For example,
the location of the peak in the invariant mass distribution of the
annihilation products can lead to mass measurements more precise
than those from cascade decays with missing energy.
The spins of the possible bound states are determined by the properties of the constituent particles. The overall production and annihilation rates are also sensitive to the color and spins of the particles.

We begin in Section~\ref{sec:top} with a discussion of bound states of the top quark. Although we will conclude that toponium is practically invisible at the LHC, this section introduces some of the formalism for studying bound states that we will
use in other sections. We proceed, in Section~\ref{sec-gogo-th} to look
at beyond-the-standard model examples.
While the same basic ideas apply to any model of new physics with
pair-produced colored particles,
to be concrete we will discuss mostly bound states of color-adjoint fermions,
such as gluinos in the minimal supersymmetric standard model (MSSM).
When a gluino pair is produced close to threshold, it can form a bound
state, called gluinonium~\cite{Goldman:1984mj,Keung:1983wz,Kuhn:1983sc}.
General features of the gluinonium production cross section and annihilation rates, including various perturbative corrections,
have been studied in~\cite{Chikovani:1996bk,BouhovaThacker:2004nh,BouhovaThacker:2006pj,Cheung:2004ad,Hagiwara:2009hq,Kauth:2009ud}.
Applying these results, we will find that annihilation decays of gluinonium are relevant for realistic SUSY spectra such as some of the SPS benchmark points~\cite{Allanach:2002nj},
as well as for more exotic scenarios such as those with a
 gluino LSP~\cite{Baer:1998pg,Raby:1998xr} or split supersymmetry~\cite{ArkaniHamed:2004fb}.
Some possibly observable final states from annihilation decays of gluinonium include dijets and $t\bar t$ pairs. In Section~\ref{sec-gogo-sim}, we study these decay modes through Monte Carlo simulation, and compare them to the relevant backgrounds.
In Section~\ref{sec-other-QCD-bs} we discuss bound states of particles with other spins, color representations and charges than the gluino. These include
squark-(anti)squark and squark-gluino bound states in the MSSM. In Section~\ref{sec-bs-new-forces} we explore what happens when
the particles are bound by a new force that is stronger than $SU(3)_\mathrm{QCD}$. Section~\ref{sec-conclusions} summarizes the various
results we find, and concludes.

\section{Wide bound states: toponium\label{sec:top}}
A reasonable place to start is a review of the bound state formalism for heavy colored particles in the standard model.
In order to study bound states using perturbation theory, the mass of the particles which bind should be larger than $\Lambda_\mathrm{QCD}$.
Bound states of charm quarks (charmonium, such as $J/\psi$) and bottom quarks (bottomonium, such as $\Upsilon$) have been well-measured
and are well-understood theoretically. Bound states of the top-quark, toponia, have been studied theoretically as well, but have
not yet been seen in colliders.
In this section, we will review what is known about toponia, and consider the prospects for their observation.
We will see that the $t\bar{t}$ invariant mass spectrum does show evidence of binding,  but
the top quark width is too large to allow for any detectable amount of annihilation decays at the LHC.
In the next sections, we will study more general non standard-model scenarios in which annihilation decays are more relevant.

\subsection{Bound state formalism}
The appropriate way to study the formation and decay of bound states is with non-relativistic
QCD (NRQCD). At high energy, relevant to the production of massive particles at colliders, QCD is a perturbative
field theory. Therefore, at leading order, the dynamics of a pair of colored particles of mass $m \gg \Lambda_{\rm QCD}$
can be described by the single-gluon exchange potential which has the form
\be
V(r) = -C\frac{\bar\alpha_s}{r} \label{V(r)} \,.
\ee
The color factor $C$ is obtained by evaluating the product of $SU(3)$ generators. Labeling the particles $1$ and $2$, this
product is
 $-T_1^a T_2^a = \frac{1}{2}((T_1^a)^2 + (T_2^a)^2-(T_1^a + T_2^a)^2)$, which leads to
\be
C = \frac{1}{2}\l(C_1 + C_2 - C_{(12)}\r) \, ,
\label{color-factor}
\ee
where $C_1$ and $C_2$ are the quadratic Casimirs for the two particles and $C_{(12)}$ that for the bound state.

The coupling constant in Eq.~\eqref{V(r)} is defined as
\be
\bar{\alpha}_s \equiv \alpha_s(a_0^{-1}) \, ,
\ee
where $a_0=(\frac{1}{2}C\bar\alpha_s m)^{-1}$ is the Bohr radius of the bound states.
This notation is chosen to emphasize that the appropriate scale at which $\alpha_s$ should be evaluated
in the Schr\"{o}dinger equation is associated with
typical momentum transfers relevant for the binding, $\mu\sim a_0^{-1}$, as opposed to the much higher energies $\mu\sim m$ which are relevant for the hard scattering production and annihilation processes. Since $\bar\alpha_s$ will eventually be raised to the third power, this can lead to a factor of $2$ or more in the overall rate. Note that the color factor $C$ depends on the color representation of the bound states and therefore there is a weak dependence in $\bar\alpha_s$ as well. Other quantities, such as the binding energy, will also depend on the representations, and to keep the notation concise, we will generally leave this dependence implicit.

When the binding particle's intrinsic width $\Gamma$ is not negligible compared to the binding energy, it must be included in the Schr\"{o}dinger
equation. We will see that this is in fact the case for the top quark.
The spectral density comprising the bound state resonances is contained in the Green's function solving
\be
\l[-\frac{\nabla^2}{m} + V(r) - \l(E + i\Gamma\r)\r] G(\vv{x},E) = \delta^{(3)}(\vv{x}) \,,
\ee
where $E=\sqrt{\hat{s}}-2m$ is the energy from the scattering event available to the system. The bound states are produced from incoming partons with momenta $\vv{k}$ which correspond to
much shorter wavelengths than the distance over which the bound state wavefunctions have support.
Therefore, the production process probes this Green's function at effectively a single point
and all we really need for the collider physics applications is $G(\vv{0},E)$. For example, by the optical theorem the partonic production cross section including bound state effects is
\be
\hat\sigma(\hat s) =
\frac{4\pi}{m^2\beta}\,\mbox{Im}\,G(\vv{0},E)\; \hat\sigma_0(\hat s) \,,
\ee
where $\beta=\sqrt{1-4m^2/\hat{s}}$ is the velocity of the two particles in the center-of-mass frame and $\hat{\sigma}_0$ is the pair production cross section without including any binding effects.
We can also use $G(\vv{0},E)$ as a propagator for the bound states, for example in computing matrix elements for the annihilation processes.

For zero angular momentum, the solution for the Green's function is~\cite{Fadin:1988fn}
\begin{multline}
G(\vv{0},E) = -\frac{m^2}{4\pi}\l[\sqrt{-\frac{E+i\Gamma}{m}} - C\bar\alpha_s\ln\l(\frac{|C|\bar\alpha_s}{2}\sqrt{-\frac{m}{E+i \Gamma}}\r) \r.\\
-\l. \frac{2}{\sqrt m}\sum_{n=1}^\infty\frac{E_n}{\sqrt{-(E+i\Gamma)} - \mbox{sign}(C)\sqrt{E_n}}\r] \,.
\label{Green-sol}
\end{multline}
Here $E_n=E_b/n^2$ are the energies of the radial excitations of the ground state whose binding energy is given by
\be
E_b = \frac{1}{4} C^2 \bar\alpha_s^2 m \,. \label{bind}
\ee
This solution encodes the resummation of Coulomb gluons to all orders in $\bar\alpha_s$. Further precision can be achieved with the inclusion of
higher order corrections to the Coulomb potential (see, e.g.,~\cite{Beneke:1999qg,Kiyo:2008bv} and references therein).

We give further details of the bound states formalism with finite width, including expressions for the production and decay rates in terms of $G(\mathbf{0},E)$,
in Appendix~\ref{sec-gen-bound}. The annihilation decays will really only be relevant for narrow width, where the formulas simplify. The formulas
in the narrow width approximation are discussed both in Appendix~\ref{sec-gen-bound} and in Section~\ref{sec-gogo-th} of the main text.

\subsection{Toponium}
For the case of a $t\bar t$ pair, like for the other quarkonia, the color decomposition
\be
\mathbf{3} \otimes \mathbf{\bar 3} = \mathbf{1} \oplus \mathbf{8}
\ee
shows that toponium can be a color singlet or a color octet. Then
\be
C_\mathbf{1} = C_F = \frac{4}{3}\,,\qqq
C_\mathbf{8} = C_F - \frac{C_A}{2} = -\frac{1}{6} \,.
\ee
Thus, the singlet has an attractive potential while the octet does not.
The top quark's intrinsic width is $\Gamma_t = 1.3$ GeV which is of the same order as its binding energy
$E_b = \frac{1}{4}C_F^2\bar\alpha_s^2 m_t = 1.5$ GeV, so we must include the width in the toponium production.

\begin{figure}
\begin{center}
\psfrag{xx}[]{$\sqrt{\hat{s}}-2 m_t$ (GeV)}
\psfrag{yy}[][][1]{$ \hat{\sigma} \times 10^9$ (GeV${}^{-2}$)}
\psfrag{A}[][][0.8]{$gg\to t\bar{t}(\mathbf{1})$}
\psfrag{B}[][][0.8]{$gg\to t\bar{t}(\mathbf{8})$}
\psfrag{C}[][][0.8]{$gg\to t\bar{t}(\mathbf{1}) \to gg$}
\psfrag{D}[][][0.8]{$\times 200$}
\psfrag{x}[]{$M$ (GeV)}
\psfrag{Y}[][][1.0]{$\frac{\rd\sigma}{\rd M}$\qq}
\psfrag{ww}[][][0.8]{\qq$(\mathrm{pb/GeV})$}
\psfrag{E}[][][0.8]{$pp\to t\bar{t} \to W^+W^-b\bar{b}$}
\psfrag{F}[][][0.8]{$(pp \to t\bar{t} \to \mathrm{dijets})\times 200$}
\psfrag{G}{}
\psfrag{H}[][][0.8]{$(pp \to t\bar{t} \to \gamma\gamma) \times 20000$}
\includegraphics[width=\textwidth]{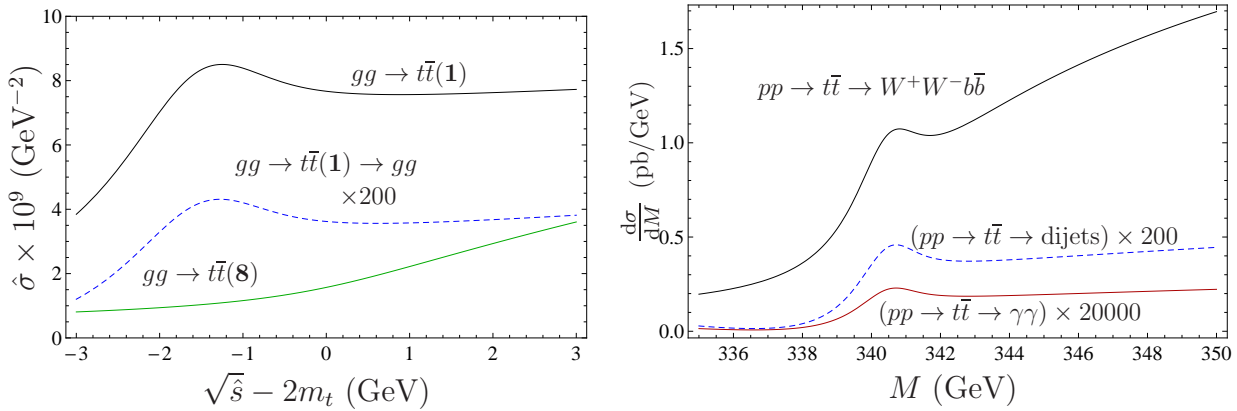}
\end{center}
\caption{Near threshold $t\bar{t}$ production. The left panel shows the partonic cross section, separated by color, with the rate
for annihilation decays scaled by 200. The right panel
shows the physical cross section for $t\bar{t}$ production at the LHC (14 TeV). The rates for
annihilation decays into dijets and $\gamma\gamma$ are shown, scaled by factors of $200$ and $20000$, respectively.}
\label{fig-toponium}
\end{figure}

The cross section for $t\bar{t}$ production in the color singlet and color octet configurations is shown in the first panel of Figure~\ref{fig-toponium}. Since the octet potential is repulsive, it has no resonance behavior. The singlet configuration shows evidence for a resonance,
but because of the large width of the top, the would-be bound state merges with the $t\bar t$ continuum. The rate for annihilation decays of the bound
state, the color singlet, is small and has to be scaled by a factor of 200 to appear on the same plot.
This can be seen analytically in the (poor) approximation of a narrow toponium, where
the rate of the annihilation process $(t\bar t) \to gg$ is~\cite{Kuhn:1987ty}
\be
\Gamma_{(t\bar t)\to gg} = \frac{64}{81}\alpha_s^2\bar\alpha_s^3m_t \simeq 4\times 10^{-3} \mbox{ GeV}
\simeq 2 \times 10^{-3} \cdot 2\Gamma_t \,.
\label{ttbar2gg}
\ee
Most of the production
cross section corresponds to processes that end up in the weak decays of the tops,
 but there are also tiny branching ratios into the various annihilation channels.
The physical cross sections times branching ratios for $t\bar{t}$ at the LHC (14 TeV)
are shown in the second panel of Figure~\ref{fig-toponium}. Higher-order QCD corrections to the production rate (but not the annihilation processes) have been studied recently in~\cite{Hagiwara:2008df,Kiyo:2008bv}, but we do not take them into account for our estimates.

In order to find evidence for binding, it is natural to search for the resonance behavior in the invariant mass of the decay products.
It is also possible to search for
the onset of the continuum contribution which looks like a step in the
invariant mass distribution. In any case, the challenge is to find a subtle feature with a width of order $\sim 1$ GeV on
top of a smooth continuum. In terms of overall rates, the best place to look for evidence of toponium might be in the reconstructed
 invariant mass spectrum from the weak decay products. However, because the reconstruction of tops is
in itself full of uncertainties, and because of the radiation broadening of the hadronic part of the top-decay final states,
this is not a promising direction to finding a subtle $1$ GeV wide feature. The dijet annihilation decay signal is similarly
hopeless.
 A slightly more promising channel, at least in terms of resolution, might have been the annihilation of toponium into photons,
which has branching ratio of $(8/9)(\alpha/\alpha_s)^2 \simeq 0.005$ relative to the annihilation into dijets~\cite{Kuhn:1987ty}.
However, for this decay
mode $\rd\sigma/\rd M \simeq 0.01$ fb/GeV which leads to too few signal events, especially compared to the standard model background which has
$\rd\sigma/\rd M \simeq 1$ fb/GeV. Similar conclusions have been reached in the limit of narrow toponium in the past~\cite{Kuhn:1992ki,Kuhn:1992qw,Fabiano:1993vx}.

\begin{figure}
\begin{center}
\psfrag{x}[]{$\sqrt{\hat{s}}-2 m_t$ (GeV)}
\psfrag{y}[][][1]{$ \hat{\sigma} \times 10^9$ (GeV${}^{-2}$)}
\psfrag{xx}[]{$\sqrt{\hat{s}}-2 m_t$ (GeV)}
\psfrag{yy}[][][1]{$ \hat{\sigma} \times10^9$ (GeV${}^{-2}$)}
\psfrag{RR}[][][0.8]{$gg\to t\bar{t}(\mathbf{1})$}
\psfrag{SS}[][][0.8]{$gg \to t\bar{t}(\mathbf{1})\to gg$}
\psfrag{TT}[][][0.8]{$\times 60$}
\includegraphics[width=\textwidth]{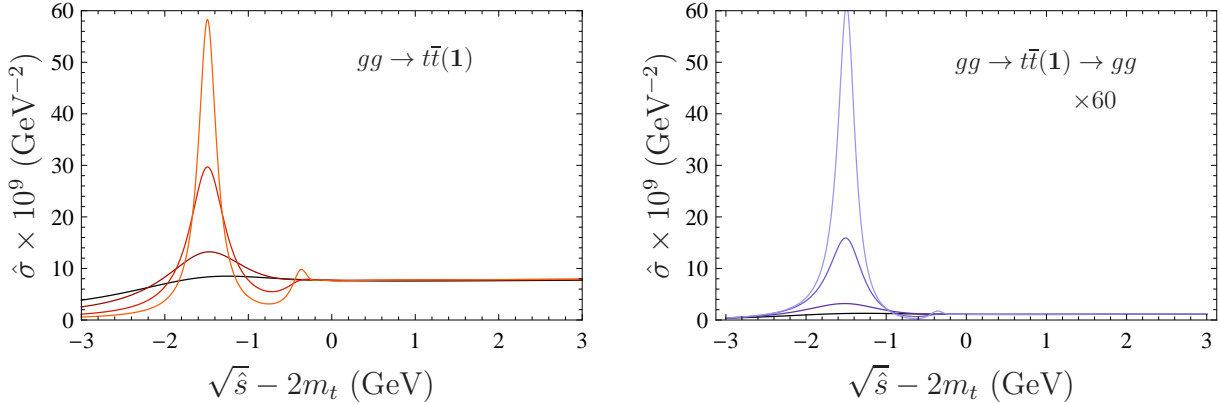}
\end{center}
\caption{Varying the top width. On the left is the $t\bar{t}$ color singlet partonic production cross section
and on the right is the rate for annihilation decays. The lowest curves correspond
to the real top, as in Figure~\ref{fig-toponium}, and the others, from bottom to top, have the top width decreased
by factors of $2,5$ and $10$. Of note are both the appearance of the bound states as well-defined resonances, and the relative
fraction of annihilation decays coming from resonance region.}
\label{fig-width}
\end{figure}

We have seen that there are two main difficulties in finding toponium.
The first is that the rate for annihilation decays is small, of order $0.2\%$ of the production
rate. The second is that the large width of the top prevents the resonance from being seen on top of the continuum background. Instead of
showing a resonance shape, the annihilation decays of toponium into dijets produce a signal which merges smoothly with the continuum.
Both of these problems would be ameliorated
if the top had a smaller intrinsic width. Figure~\ref{fig-width} shows what happens to the production and annihilation decay rates
as the top quark width $\Gamma_t$ is decreased. First note from the first panel that the resonance becomes much sharper for smaller width.
Not only does the bound state become well-defined but its first radial excitations become visible too.
 Thus the distinction from the continuum in any decay mode will be sharper. Second, note from the second plot the relative importance of the annihilation decays of the resonance compared to the continuum.
Of course, the top width is fixed, and known. But there are beyond the standard model scenarios which have phenomenology similar to a narrow top.
One example is bound states of gluinos, gluinonium, to which we now turn.

\section{Narrow bound states: Gluinonium \label{sec-gogo-th}}
Having found that toponium is practically impossible to see at the LHC, let us consider in what situation bound states of colored particles might be detectable.
We saw in Figure~\ref{fig-width} that the strength of the annihilation decay signal is enhanced when
the single particle width is decreased.
If we can neglect the particle's intrinsic width in the bound state formalism, the
decay of a resonance can be simply expressed in terms of its wavefunction $\psi(\vv{x})$. The relevant rates are then
\be
\Gamma_{\rm ann} \sim \frac{\alpha_s^2}{m^2}\l|\psi(\vv{0})\r|^2 \sim \alpha_s^2\bar\alpha_s^3 m \,, \qqq
E_b \sim \bar\alpha_s^2 m \,.
\label{scales}
\ee
So $\Gamma_{\rm ann}/E_b \sim \alpha_s^2\bar\alpha_s \ll 1$.
Thus, bound states in this situation will be very narrow compared to the binding energy, like in Figure~\ref{fig-width}.

As a canonical example, we will now consider bound states of gluinos, gluinonia. We
will return to other cases in Sections~\ref{sec-other-QCD-bs} and~\ref{sec-bs-new-forces}.
Unlike the top whose properties we know, the gluino has, minimally, two free parameters: its mass $m_\go$ and its width $\Gamma_\go$. In many scenarios the gluino decays to a squark and a quark, with a decay rate scaling like
\be
\Gamma_\go \sim \alpha_s m_\go \,.
\label{strong-go-decay}
\ee
This leads to $\Gamma_\go \sim E_b$ and the gluinonium will be broad like the toponium.
However, if these two-body decays are kinematically forbidden, which happens
when all of the squarks are heavier than the gluino, the gluino can only decay through off-shell intermediate states (see~\cite{Barnett:1987kn,Toharia:2005gm} for detailed studies of the various possible decay modes in this situation).
In this case, the decay rate is significantly suppressed. For example, if the dominant decay mode is to quarks and a chargino or neutralino through an off-shell squark, the rate will scale like
\begin{equation}
\Gamma_\go
\sim
\frac{\alpha_{\mathrm{em}} \alpha_s m_\go}{16\pi\sin^2\theta_W}\left(\frac{m_\go}{m_\sq}\right)^4 \,.
\label{weak-go-decay}
\end{equation}
In principle, this can be as small as one likes if the squarks are heavy, which enhances the fraction of annihilation decays. Since the rate goes as the
fourth power of $m_\go/m_\sq$, the squarks only need to be slightly heavier than the gluinos to saturate the rate for annihilation decays.

Keep in mind, for the annihilation decays to occur, the gluinos have to bind, which only happens near threshold.
If the gluinos are produced with relativistic speeds, they still undergo single-particle decays rather than annihilating.
In the case where the squarks are completely decoupled like in split supersymmetry~\cite{ArkaniHamed:2004fb},
the relativistic gluinos may even form bound states with standard model particles, for example, forming R-hadrons (see, e.g.,~\cite{Baer:1998pg,Raby:1998xr,Kilian:2004uj,Hewett:2004nw}).
Such exotic phenomenology is largely irrelevant to the cross section for annihilation decays in the resonance region, which is the place to look for gluinonia.

\subsection{Narrow gluinonia}
Since we will be interested in the limit $\Gamma_\go, \Gamma_{\rm ann} \ll E_b$, we can use the narrow width approximation to simplify the bound-state formalism.
In the narrow width limit, the production and annihilation of the bound states are determined by finding
the energy eigenstates of the Schr\"{o}dinger equation from the potential~\eqref{V(r)}.
The solutions are then wavefunctions $\psi(\vv{x})$, analogous to the wavefunctions of the hydrogen atom. For $S$-waves,
these are simply functions of the radial coordinate: $\psi_n(\vv{x}) = \frac{1}{2\sqrt\pi}\,R_n(r)$, where $n$ is the radial quantum number. In terms of the formalism of Section~\ref{sec:top}, below
the kinematic pair-production threshold, the imaginary part of the Green's function reduces to
\be
\mbox{Im}\,G(\vv{0}) = \pi\sum_n \l|\psi_n(\vv{0})\r|^2\,\delta(E+E_n) \,.
\ee
The short-distance part of the bound state production process is determined by the same matrix element ${\cal M}_0$ that describes the production of the constituent particles in the continuum.
The matrix element for binding is given by~\cite{Peskin-Schroeder,Kuhn:1979bb}
\be
\cM_{\rm bound} = \sqrt{\frac{2}{M}}\,\psi(\vv{0})\,\cM_0 \,,
\label{times-wavefunction}
\ee
where $M \simeq 2m$ is the mass of the bound state. Note that, although the continuum production rate vanishes at threshold $\sqrt{\hat{s}}=2m$,
this is due to phase space;  the matrix element $\cM_0$ at threshold is finite.

For identical particles, a factor of $1/2$ needs to be added in (\ref{times-wavefunction}) and the wave function needs
to satisfy an appropriate (anti)symmetry condition. For an $S$-wave spin-$0$ color-symmetric (or spin-$1$ color-antisymmetric) bound state of two gluinos (identical fermions), we should take the spatial part as
\be
\psi_{\go\go}(\vv{x}) = {1\over\sqrt2}\l[\psi(\vv{x}) + \psi(\vv{-x})\r] = \sqrt{2}\,\psi(\vv{x}) \,.
\label{identical-factor}
\ee
These factors result in an overall factor of $1/2$ in the bound state production and annihilation rates. Following the convention in the literature, $\psi(\vv{x})$ in our expressions will still be normalized as in the hydrogen atom, while the extra factor of $1/2$ will be included explicitly in the prefactor.

The binding energy, Bohr radius, and the wavefunction at $\vv{x}=0$ for
the ground state are given by
\be\label{hydrogen-like}
E_b = \frac{C^2\bar \alpha_s^2 m}{4}, \qq
a_0 = \frac{2}{C\bar \alpha_s m},\qq
\l|\psi(\vv{0})\r|^2 = \frac{1}{\pi a_0^3} = \frac{C^3\bar \alpha_s^3 m^3}{8\pi} \,.
\ee
As before, we have made a notational distinction for the
strong coupling constant evaluated at scales relevant for the binding $\bar\alpha_s\equiv\alpha_s(a_0^{-1})$
and the strong coupling scale relevant for the production, $\alpha_s=\alpha_s(m)$.
To get a feel for the sizes of the quantities in (\ref{hydrogen-like}), we can try $C=3$ and $m=300$ GeV, giving
$E_b \simeq 11$ GeV, $1/a_0 \simeq 56$ GeV and $\l|\psi(\vv{0})\r|^2 \simeq  (40 \, \mathrm{GeV})^3$.

Excited states are much less important than the ground state.
The radial excitations have wavefunctions which
at $\vv{x}=0$ are given by $\psi_n(\vv{0}) = \psi(\vv{0})/n^{3/2}$,
so their decay rates and cross sections are
suppressed by $1/n^3$. All angular-excited states have $\psi(\vv{0}) = 0$, so their rates depend on the derivative of $\psi(\vv{x})$ with respect to the radial coordinate. These rates are then proportional to $|\psi'(\vv{0})|^2/m^2 \sim \bar\alpha_s^5 m^3$ rather than $|\psi(\vv{0})|^2 \sim \bar\alpha_s^3 m^3$~\cite{Kuhn:1979bb}, which suppresses their decay rates and production cross sections by a factor of $\bar\alpha_s^2$ compared to the ground states. For this reason, we will only consider $S$-wave gluinonia. However $P$-wave gluinonia have been mentioned in the context of diffractive processes in~\cite{Khoze:2001xm,Bussey:2006vx}. Note also that because their annihilation rates are suppressed, they can be corrected at the leading order by subleading terms in NRQCD. For example, in quarkonia, $P$ states can have small admixtures of states that are made of a $q\bar q$ pair in an $S$ color-octet state + a gluon~\cite{Bodwin:1992ye}.

The color representation of a pair of gluinos can be one of the following:
\be
\mathbf{8} \otimes \mathbf{8} = \mathbf{1_S} \oplus \mathbf{8_S} \oplus \mathbf{8_A} \oplus \mathbf{10_A} \oplus \mathbf{\bar{10}_A} \oplus \mathbf{27_S} \,.
\ee
Here $\mathbf{S}$ and $\mathbf{A}$ subscripts refer to whether the wavefunction is symmetric or anti-symmetric in color.
Based on (\ref{color-factor}), the color factors are
\be\label{gluinonium-color-factors}
C_\mathbf{1} = C_A=3\,,\qq
C_\mathbf{8_S} = C_\mathbf{8_A} =\frac{1}{2} C_A=\frac{3}{2}\,,\qq
C_\mathbf{10} = C_\mathbf{\bar{10}} = 0\,,\qq
C_\mathbf{27} = -1 \,,
\ee
so the potential is attractive for $\mathbf{1}$, $\mathbf{8_S}$
and $\mathbf{8_A}$, zero for $\mathbf{10}$ and $\mathbf{\bar{10}}$,
which are not produced at leading order~\cite{Kulesza:2009kq},
and repulsive for $\mathbf{27}$. Note that the color factors are raised to the third power in the expression for $|\psi(\vv{0})|^2$ in (\ref{hydrogen-like}),
and this can make a large difference in the cross section. It enhances the color singlet production over the color octet by
a factor of 8.
$S$-wave bound states exist for all the three attractive representations. Since the gluinos are identical spin-$\frac{1}{2}$ Majorana fermions, the color-symmetric states must be pseudoscalars ($J^{PC} = 0^{-+}$)
\be
\l|^1S_0(\mathbf{1})\r\rangle = \frac{1}{\sqrt 8}\,\delta_{ab}\,, \qqq
\l|^1S_0(\mathbf{8_S})\r\rangle_c = \sqrt{\frac{3}{5}}\,d_{abc}
\label{wfs}
\ee
and the color-antisymmetric state must be a vector ($J^{PC} = 1^{--}$)
\be
\l|^3S_1(\mathbf{8_A})\r\rangle_c = \frac{1}{\sqrt 3}\,f_{abc} \,,
\ee
where $^{2S+1}L_J$ describes the rotational quantum numbers.

\subsection{Annihilation decays}
\begin{figure}
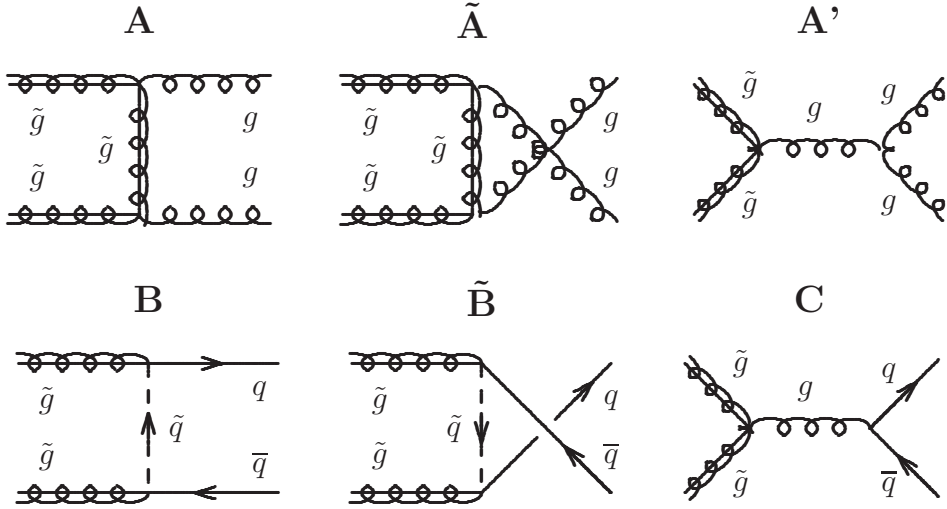

$$\beginpicture
\setcoordinatesystem units <0.7\tdim,0.7\tdim>
\stpltsmbl
\put {\large\textbf{A}} at 0 70
\plot -70 35 0 35 0 -35 -70 -35 /
\ellipticalarc axes ratio 2:1 150 degrees from -62 31 center at -69 35
\ellipticalarc axes ratio 2:1 280 degrees from -47 31 center at -54 35
\ellipticalarc axes ratio 2:1 280 degrees from -32 31 center at -39 35
\ellipticalarc axes ratio 2:1 280 degrees from -17 31 center at -24 35
\ellipticalarc axes ratio 2:1 220 degrees from -0 35 center at -10 35
\ellipticalarc axes ratio 2:1 -220 degrees from 0 35 center at 10 35
\ellipticalarc axes ratio 2:1 -280 degrees from 17 31 center at 24 35
\ellipticalarc axes ratio 2:1 -280 degrees from 32 31 center at 39 35
\ellipticalarc axes ratio 2:1 -280 degrees from 47 31 center at 54 35
\ellipticalarc axes ratio 2:1 -150 degrees from 62 31 center at 69 35
\ellipticalarc axes ratio 2:1 -150 degrees from -62 -31 center at -69 -35
\ellipticalarc axes ratio 2:1 -280 degrees from -47 -31 center at -54 -35
\ellipticalarc axes ratio 2:1 -280 degrees from -32 -31 center at -39 -35
\ellipticalarc axes ratio 2:1 -280 degrees from -17 -31 center at -24 -35
\ellipticalarc axes ratio 2:1 -220 degrees from 0 -35 center at -10 -35
\ellipticalarc axes ratio 2:1 220 degrees from 0 -35 center at 10 -35
\ellipticalarc axes ratio 2:1 280 degrees from 17 -31 center at 24 -35
\ellipticalarc axes ratio 2:1 280 degrees from 32 -31 center at 39 -35
\ellipticalarc axes ratio 2:1 280 degrees from 47 -31 center at 54 -35
\ellipticalarc axes ratio 2:1 150 degrees from 62 -31 center at 69 -35
\startrotation by 0 -1 about 0 30
\ellipticalarc axes ratio 2:1 -220 degrees from -5 29 center at  5 29
\ellipticalarc axes ratio 2:1 -280 degrees from 12 25 center at 19 29
\ellipticalarc axes ratio 2:1 -280 degrees from 27 25 center at 34 29
\ellipticalarc axes ratio 2:1 -280 degrees from 42 25 center at 49 29
\ellipticalarc axes ratio 2:1 -180 degrees from 57 25 center at 64 29
\stoprotation
\put {$\go$} at -55 15
\put {$\go$} at -55 -15
\put {$\go$} at -18 0
\put {$g$} at 60 15
\put {$g$} at 60 -15
\linethickness=0pt
\putrule from 0 -60 to 0 80
\putrule from -90 0 to 90 0
\endpicture
\beginpicture
\setcoordinatesystem units <0.7\tdim,0.7\tdim>
\stpltsmbl
\put {\large\textbf{\~{A}}} at 0 70
\plot -70 35 0 35 0 -35 -70 -35 /
\ellipticalarc axes ratio 2:1 150 degrees from -62 31 center at -69 35
\ellipticalarc axes ratio 2:1 280 degrees from -47 31 center at -54 35
\ellipticalarc axes ratio 2:1 280 degrees from -32 31 center at -39 35
\ellipticalarc axes ratio 2:1 280 degrees from -17 31 center at -24 35
\ellipticalarc axes ratio 2:1 220 degrees from -0 35 center at -10 35
\ellipticalarc axes ratio 2:1 -150 degrees from -62 -31 center at -69 -35
\ellipticalarc axes ratio 2:1 -280 degrees from -47 -31 center at -54 -35
\ellipticalarc axes ratio 2:1 -280 degrees from -32 -31 center at -39 -35
\ellipticalarc axes ratio 2:1 -280 degrees from -17 -31 center at -24 -35
\ellipticalarc axes ratio 2:1 -220 degrees from 0 -35 center at -10 -35
\startrotation by 0 -1 about 0 30
\ellipticalarc axes ratio 2:1 -220 degrees from -5 29 center at  5 29
\ellipticalarc axes ratio 2:1 -280 degrees from 12 25 center at 19 29
\ellipticalarc axes ratio 2:1 -280 degrees from 27 25 center at 34 29
\ellipticalarc axes ratio 2:1 -280 degrees from 42 25 center at 49 29
\ellipticalarc axes ratio 2:1 -150 degrees from 57 25 center at 64 29
\stoprotation
\startrotation by 0.7 -0.75 about 0 30
\ellipticalarc axes ratio 2:1 -220 degrees from 0 35 center at 10 35
\ellipticalarc axes ratio 2:1 -280 degrees from 17 31 center at 24 35
\ellipticalarc axes ratio 2:1 -280 degrees from 32 31 center at 39 35
\ellipticalarc axes ratio 2:1 -280 degrees from 47 31 center at 54 35
\ellipticalarc axes ratio 2:1 -280 degrees from 62 31 center at 69 35
\ellipticalarc axes ratio 2:1 -280 degrees from 77 31 center at 84 35
\ellipticalarc axes ratio 2:1 -150 degrees from 92 31 center at 99 35
\stoprotation
\startrotation by 0.7 0.75 about 0 -30
\ellipticalarc axes ratio 2:1 220 degrees from 0 -35 center at 10 -35
\ellipticalarc axes ratio 2:1 280 degrees from 17 -31 center at 24 -35
\ellipticalarc axes ratio 2:1 280 degrees from 32 -31 center at 39 -35
\ellipticalarc axes ratio 2:1 280 degrees from 47 -31 center at 54 -35
\ellipticalarc axes ratio 2:1 280 degrees from 62 -31 center at 69 -35
\ellipticalarc axes ratio 2:1 280 degrees from 77 -31 center at 84 -35
\ellipticalarc axes ratio 2:1 150 degrees from 92 -31 center at 99 -35
\stoprotation
\put {$\go$} at -55 15

\put {$\go$} at -55 -15
\put {$\go$} at -18 0
\put {$g$} at 75 15
\put {$g$} at 75 -15
\linethickness=0pt
\putrule from 0 -60 to 0 80
\putrule from -90 0 to 90 0
\endpicture
\beginpicture
\setcoordinatesystem units <0.7\tdim,0.7\tdim>
\stpltsmbl
\put {\large\textbf{A'}} at -33 70
\plot -100 35 -65 0 /
\startrotation by 0.6 -0.6 about -100 35
\ellipticalarc axes ratio 2:1 150 degrees from -92 31 center at -99 35
\ellipticalarc axes ratio 2:1 280 degrees from -77 31 center at -84 35
\ellipticalarc axes ratio 2:1 280 degrees from -62 31 center at -69 35
\ellipticalarc axes ratio 2:1 280 degrees from -47 31 center at -54 35
\stoprotation
\plot -100 -35 -65 0 /
\startrotation by 0.6 0.6 about -100 -35
\ellipticalarc axes ratio 2:1 -150 degrees from -92 -31 center at -99 -35
\ellipticalarc axes ratio 2:1 -280 degrees from -77 -31 center at -84 -35
\ellipticalarc axes ratio 2:1 -280 degrees from -62 -31 center at -69 -35
\ellipticalarc axes ratio 2:1 -280 degrees from -47 -31 center at -54 -35
\stoprotation
\ellipticalarc axes ratio 2:1 220 degrees from  0  0 center at -10 0
\ellipticalarc axes ratio 2:1 280 degrees from -17 -4 center at -24 0
\ellipticalarc axes ratio 2:1 280 degrees from -32 -4 center at -39 0
\ellipticalarc axes ratio 2:1 260 degrees from -47 -4 center at -54 0
\startrotation by 0.6 0.6 about 37 35
\ellipticalarc axes ratio 2:1 -150 degrees from  29 31 center at  36 35
\ellipticalarc axes ratio 2:1 -280 degrees from  14 31 center at  21 35
\ellipticalarc axes ratio 2:1 -280 degrees from  -1 31 center at   6 35
\ellipticalarc axes ratio 2:1 -280 degrees from -16 31 center at  -9 35
\stoprotation
\startrotation by 0.6 -0.6 about 37 -35
\ellipticalarc axes ratio 2:1 150 degrees from  29 -31 center at  36 -35
\ellipticalarc axes ratio 2:1 280 degrees from  14 -31 center at  21 -35
\ellipticalarc axes ratio 2:1 280 degrees from  -1 -31 center at   6 -35
\ellipticalarc axes ratio 2:1 280 degrees from -16 -31 center at  -9 -35
\stoprotation
\put {$\go$} at -70 35
\put {$\go$} at -70 -30
\put {$g$} at -35 20
\put {$g$} at 5 30
\put {$g$} at 5 -30
\linethickness=0pt
\putrule from 0 -60 to 0 80
\putrule from -130 0 to 90 0
\endpicture
$$
$$
\beginpicture
\setcoordinatesystem units <0.7\tdim,0.7\tdim>
\stpltsmbl
\put {\large\textbf{B}} at 0 70
\plot -70 35 70 35 /
\plot -70 -35 70 -35 /
\barrow from 5 35 to 40 35
\barrow from 70 -35 to 25 -35
\ellipticalarc axes ratio 2:1 150 degrees from -62 31 center at -69 35
\ellipticalarc axes ratio 2:1 280 degrees from -47 31 center at -54 35
\ellipticalarc axes ratio 2:1 280 degrees from -32 31 center at -39 35
\ellipticalarc axes ratio 2:1 280 degrees from -17 31 center at -24 35
\ellipticalarc axes ratio 2:1 220 degrees from -0 35 center at -10 35
\ellipticalarc axes ratio 2:1 -150 degrees from -62 -31 center at -69 -35
\ellipticalarc axes ratio 2:1 -280 degrees from -47 -31 center at -54 -35
\ellipticalarc axes ratio 2:1 -280 degrees from -32 -31 center at -39 -35
\ellipticalarc axes ratio 2:1 -280 degrees from -17 -31 center at -24 -35
\ellipticalarc axes ratio 2:1 -220 degrees from 0 -35 center at -10 -35
\setdashes
\plot 0 35 0 -35 /
\setsolid
\barrow from 0 -8 to 0 8
\put {$\go$} at -55 15
\put {$\go$} at -55 -15
\put {$\sq$} at 15 0
\put {$q$} at 60 20
\put {$\bar q$} at 60 -20
\linethickness=0pt
\putrule from 0 -60 to 0 80
\putrule from -90 0 to 90 0
\endpicture
\beginpicture
\setcoordinatesystem units <0.7\tdim,0.7\tdim>
\stpltsmbl
\put {\large\textbf{\~{B}}} at 0 70
\plot -70 35 0 35 70 -35 /
\plot -70 -35 0 -35 30 -5 /
\plot 40 5 70 35 /
\barrow from 70 -35 to 45 -10
\barrow from 50 15 to 60 25
\ellipticalarc axes ratio 2:1 150 degrees from -62 31 center at -69 35
\ellipticalarc axes ratio 2:1 280 degrees from -47 31 center at -54 35
\ellipticalarc axes ratio 2:1 280 degrees from -32 31 center at -39 35
\ellipticalarc axes ratio 2:1 280 degrees from -17 31 center at -24 35
\ellipticalarc axes ratio 2:1 220 degrees from -0 35 center at -10 35
\ellipticalarc axes ratio 2:1 -150 degrees from -62 -31 center at -69 -35
\ellipticalarc axes ratio 2:1 -280 degrees from -47 -31 center at -54 -35
\ellipticalarc axes ratio 2:1 -280 degrees from -32 -31 center at -39 -35
\ellipticalarc axes ratio 2:1 -280 degrees from -17 -31 center at -24 -35
\ellipticalarc axes ratio 2:1 -220 degrees from 0 -35 center at -10 -35
\setdashes
\plot 0 35 0 -35 /
\setsolid
\barrow from 0 8 to 0 -8
\put {$\go$} at -55 15
\put {$\go$} at -55 -15
\put {$\sq$} at -15 0
\put {$q$} at 70 15
\put {$\bar q$} at 70 -15
\linethickness=0pt
\putrule from 0 -60 to 0 80
\putrule from -90 0 to 90 0
\endpicture
\beginpicture
\setcoordinatesystem units <0.7\tdim,0.7\tdim>
\stpltsmbl
\put {\large\textbf{C}} at -33 70
\plot -100 35 -65 0 /
\startrotation by 0.6 -0.6 about -100 35
\ellipticalarc axes ratio 2:1 150 degrees from -92 31 center at -99 35
\ellipticalarc axes ratio 2:1 280 degrees from -77 31 center at -84 35
\ellipticalarc axes ratio 2:1 280 degrees from -62 31 center at -69 35
\ellipticalarc axes ratio 2:1 280 degrees from -47 31 center at -54 35
\stoprotation
\plot -100 -35 -65 0 /
\startrotation by 0.6 0.6 about -100 -35
\ellipticalarc axes ratio 2:1 -150 degrees from -92 -31 center at -99 -35
\ellipticalarc axes ratio 2:1 -280 degrees from -77 -31 center at -84 -35
\ellipticalarc axes ratio 2:1 -280 degrees from -62 -31 center at -69 -35
\ellipticalarc axes ratio 2:1 -280 degrees from -47 -31 center at -54 -35
\stoprotation
\setsolid
\ellipticalarc axes ratio 2:1 220 degrees from  0  0 center at -10 0
\ellipticalarc axes ratio 2:1 280 degrees from -17 -4 center at -24 0
\ellipticalarc axes ratio 2:1 280 degrees from -32 -4 center at -39 0
\ellipticalarc axes ratio 2:1 260 degrees from -47 -4 center at -54 0
\plot 37 37 0 0 37 -37 /
\barrow from 0 0 to 25 25
\barrow from 35 -35 to 15 -15
\put {$\go$} at -70 35
\put {$\go$} at -70 -30
\put {$g$} at -35 20
\put {$q$} at 10 30
\put {$\bar q$} at 10 -30
\linethickness=0pt
\putrule from 0 -60 to 0 80
\putrule from -120 0 to 90 0
\endpicture$$
\caption{Diagrams for annihilation decays of gluinonia.}
\label{fig-gogo-annih}
\end{figure}

The gluinonia with different spins will have different annihilation modes, so let us consider the pseudoscalar and vector bound states separately.
Pseudoscalar gluinonia in either the color singlet, $\mathbf{1}$,
or adjoint, $\mathbf{8_S}$, representation will annihilate predominantly into two gluons via the diagrams A and \~{A} in Figure~\ref{fig-gogo-annih}. Including also the normalizations from (\ref{times-wavefunction}), (\ref{identical-factor}) and (\ref{wfs}), the resulting rates are~\cite{Goldman:1984mj,Keung:1983wz,Cheung:2004ad}
\be
\Gamma(\mathbf{1}\to gg) = \frac{18\pi\alpha_s^2}{m_\go^2}\l|\psi_\mathbf{1}(\vv{0})\r|^2 = \frac{243}{4} \alpha_s^2\bar\alpha_s^3 m_\go \label{1annih} \,,
\ee
and~\cite{Cheung:2004ad}\footnote{The rate (\ref{8Sannih}) is $8$ times smaller than that found by~\cite{Goldman:1984mj} which was used in~\cite{Chikovani:1996bk,BouhovaThacker:2004nh,BouhovaThacker:2006pj}. The expression of~\cite{Cheung:2004ad} is the correct one, as can be seen by comparing the $\mathbf{1}$ and $\mathbf{8_S}$ cases. For $\mathbf{1}\to gg$, diagrams A and \~{A} from Figure~\ref{fig-gogo-annih} include the color factor
$$
\frac{1}{\sqrt 8}\delta_{ab}f_{age}f_{ehb} = -\frac{3}{\sqrt 8}\,\delta_{gh}
$$
where $g$ and $h$ are the adjoint indices of the gluons, while $\mathbf{8_S} \to gg$ has the color factor
$$
\sqrt{\frac{3}{5}}d_{abc}f_{age}f_{ehb} = -\sqrt{\frac{3}{5}}\l(d_{gac}f_{bea} + d_{gba}f_{cea}\r)f_{ehb} = -\sqrt{\frac{3}{5}}\frac{3}{2}\,d_{ghc}
$$
Squaring and summing over $g$ and $h$, we obtain $\Gamma_{\mathbf{1}\to gg}/\Gamma_{\mathbf{8_S}\to gg} = 4\,\l|\psi_\mathbf{1}(\vv{0})\r|^2/\l|\psi_\mathbf{8_S}(\vv{0})\r|^2$, as in~\cite{Cheung:2004ad}. The mistake of~\cite{Goldman:1984mj} is that they summed over the color index $c$ of the decaying gluinonium in their (2.10).\label{error-explanation}}

\be
\Gamma(\mathbf{8_S}\to gg) = \frac{9\pi\alpha_s^2}{2m_\go^2}\l|\psi_\mathbf{8_S}(\vv{0})\r|^2 = \frac{243}{128} \alpha_s^2\bar\alpha_s^3 m_\go \label{8Sannih}\,.
\ee

The pseudoscalar gluinonia can also decay into quarks via diagrams B and \~{B} in Figure~\ref{fig-gogo-annih}. The annihilation decay to light $q\bar q$ pairs is suppressed by the quark masses by chirality considerations.
However, decays to $t\bar t$ may have a reasonable rate~\cite{BouhovaThacker:2004nh,Kauth:2009ud}. The branching ratio for the $\mathbf{1}$ gluinonium is~\cite{Kauth:2009ud}
\be
\frac{\Gamma(\mathbf{1}\to t\bar t)}{\Gamma(\mathbf{1}\to gg)}
= \frac{16}{27}\,\sqrt{1 - \frac{m_t^2}{m_\go^2}}\,\frac{m_t^2 m_\go^2}{\l(m_\go^2 + m_\st^2 - m_t^2\r)^2}
\label{1annih-ttbar}
\ee
assuming that both stops have the same mass $m_\st$. For the $\mathbf{8_S}$ gluinonium we similarly find
\be
\frac{\Gamma(\mathbf{8_S}\to t\bar t)}{\Gamma(\mathbf{8_S}\to gg)}
= \frac{20}{27}\,\sqrt{1 - \frac{m_t^2}{m_\go^2}}\,\frac{m_t^2 m_\go^2}{\l(m_\go^2 + m_\st^2 - m_t^2\r)^2} \,.
\label{8Sannih-ttbar}
\ee
For $m_\sq \sim m_\go$, these branching ratios are of the order of $5\%$ for a 300 GeV gluino and $0.5\%$ for a $1$ TeV gluino. There are no other decay modes of the pseudoscalars with considerable branching fractions.

The vector color-octet ($\mathbf{8_A}$) gluinonium
is not produced by gluon fusion at tree level and does not decay to
gluons because the sum of diagrams A, \~{A} and A' vanishes at threshold. It annihilates into $q\bar q$ pairs via diagrams B, \~{B} and C in Figure~\ref{fig-gogo-annih}. The rate is (based on~\cite{Cheung:2004ad} and the $m_\sq$-dependent prefactor taken from, e.g., the calculation of~\cite{Beenakker:1996ch} for $q\bar q \to \go\go$)
\be
\Gamma(\mathbf{8_A}\to q\bar q) =
\sum_\sq\left(\frac{m_\sq^2 - m_\go^2}{m_\sq^2 + m_\go^2}\right)^2
\frac{\pi\alpha_s^2}{2m_\go^2}\l|\psi_\mathbf{8_A}(\vv{0})\r|^2 =
\sum_\sq\left(\frac{m_\sq^2 - m_\go^2}{m_\sq^2 + m_\go^2}\right)^2
\frac{27}{128}\alpha_s^2\bar\alpha_s^3 m_\go \,,
\label{8Aannihfull}
\ee
where the summation is over the flavors and chiralities of the squarks.

Interestingly, whenever a squark has $m_\sq = m_\go$ its contribution vanishes  because of the destructive interference of diagrams B and \~{B}, which depend on the squark masses $m_\sq$, with diagram C, which does not. If the squarks are much heavier than the gluinos, only diagram C contributes, and
(\ref{8Aannihfull}) reduces to~\cite{Cheung:2004ad}
\be
\Gamma(\mathbf{8_A}\to q\bar q) =
\frac{27}{64}n_f\alpha_s^2\bar\alpha_s^3 m_\go \,,
\label{8Aannih}
\ee
where $n_f$ is the number of quark flavors.

Numerically, all the annihilation widths are very small, below $1$ GeV. Since these decay widths are many orders of magnitude smaller than the gluinonium mass, the gluinonia can effectively be treated as 0-width resonances decaying to dijets and $t\bar{t}$. The annihilation decay modes then have the potential to lead to a precise measurement of the gluino mass. If the decay rates are small enough
the colored gluinonia will even hadronize with standard model particles before annihilating. However, since $1/a_0 \gg \Lambda_{\rm{QCD}}$, the gluinonium within the hadron will be much smaller than the hadron itself, and so the annihilation decays are insensitive to whether or not the gluinonia are confined.

As expected from (\ref{scales}), in cases when the gluinonium
decays primarily by annihilation ($\Gamma^{\rm ann} \gg 2\Gamma_\go$), its width is much smaller than the binding energy.
In fact, for the three bound states,
\be
\frac{1}{E_b}\Gamma^{\rm ann}(\mathbf{1},\mathbf{8_S},\mathbf{8_A})
\sim \alpha_s^2\bar\alpha_s (27,\,\frac{27}{8},\,\frac{3}{4}n_f)\,
\sim (0.03, 0.003, 0.004) \,.
\ee
Such gluinonia will appear as distinct states separated from the gluino pair continuum. This also provides a self-consistency check on our use of the narrow width approximation instead of the full NRQCD formalism which we needed for toponium.

\subsection{Cross sections at the LHC\label{sec-gluinonium-prod}}
As we observed, annihilation decays dominate over other gluinonia decay modes when
the squarks are all at least as heavy as the gluino.
This is of course possible in parts of the MSSM, since all the superpartner masses are free parameters.
To see whether this parameter region corresponds to any viable models motivated by other considerations, we
list in   Table~\ref{tab-gogo-SPS} the relevant
parameters for a standard set of MSSM benchmark points, the SPS points~\cite{Allanach:2002nj}. In many of these scenarios $2\Gamma_\go \gg \Gamma^{\rm ann}$ so the gluinos will
decay much more than annihilate, or, if $2\Gamma_\go \gtrsim E_b$, they will even decay before forming bound states, reducing the situation to one similar to toponium. These decays will contribute to the usual single-gluino decay signature as in the continuum gluino-pair production and the size of this contribution has been studied recently in~\cite{Hagiwara:2009hq}. We are interested in cases where $2\Gamma_\go \lesssim \Gamma^{\rm ann} \ll E_b$ since this is when there is a chance to see the annihilation decays. For the pseudoscalar gluinonia ($\mathbf{1}$ and $\mathbf{8_S}$) the rate for annihilation decays is significant for the SPS 2 (mSUGRA focus point) and SPS 8 (GMSB with neutralino NLSP) scenarios. In these models, two body decays of the gluino are forbidden. For the other SPS points, the gluino decays too fast for annihilation decays to be relevant. The annihilation rate of the vector gluinonium ($\mathbf{8_A}$) turns out to typically have a sizeable suppression due to the $m_\sq$-dependent factor in (\ref{8Aannihfull}), as also shown in Table~\ref{tab-gogo-SPS}. Nevertheless, its annihilation is still significant at SPS 2. We can conclude that there are models in which annihilation decays may be observable. Since including the effect of a finite branching ratio is trivial, in our simulations we will assume $\Gamma_\go = 0$. It is this limit that is indicated by the last rows in Table~\ref{tab-gogo-SPS}. We consider two representative cases: all the squarks degenerate with the gluinos (which is roughly the case in most of the SPS points) and all the squarks much heavier than the gluinos (like in split SUSY).

\begin{table}[t]
$$\begin{array}{|c|c|c|c|c|c|c|c|c|c|c|c|c|c|}
\hline
\,\mbox{Model}
\,&\, m_\go
\,&\, m_\st
\,&\, E_b^\mathbf{1}
\,&\, E_b^\mathbf{8}
\,&\, 2\,\Gamma_\go
\,&\, \Gamma_\mathbf{1}^{\rm ann}
\,&\, \Gamma_\mathbf{8_S}^{\rm ann}
\,&\, \l(\Gamma_\mathbf{8_A}^{\rm ann}\r)
\,&\, \Gamma_\mathbf{8_A}^{\rm ann}
\,&\, \sigma^{\rm cont}
\,&\, \sigma^{\rm bound}
\,&\, \sigma^{\rm ann}_{jj}
\,&\, \sigma^{\rm ann}_{t\bar t}
\\\,
\,&\, \mbox{GeV}
\,&\, \mbox{GeV}
\,&\, \mbox{GeV}
\,&\, \mbox{GeV}
\,&\, \mbox{GeV}
\,&\, \mbox{MeV}
\,&\, \mbox{MeV}
\,&\, \mbox{MeV}
\,&\, \mbox{MeV}
\,&\, \mbox{pb}
\,&\, \mbox{fb}
\,&\, \mbox{fb}
\,&\, \mbox{fb}
\\\hline
\mbox{SPS 1a}&  607 & 400 & 18 & 5.3 & 11     & 460 & 18 & 25 & 0.5 & 6.3   & 240   & 7.2   & 0.14 \\
\mbox{SPS 1b}&  938 & 660 & 25 & 7.4 & 20     & 530 & 21 & 29 & 0.6 & 0.30  & 13    & 0.3   & 0.00 \\
\mbox{SPS 2} &  782 & 950 & 22 & 6.5 & 0.0052 & 500 & 20 & 27 & 8.2 & 1.7   & 51    & 46    & 0.41 \\
\mbox{SPS 3} &  935 & 648 & 25 & 7.4 & 23     & 530 & 21 & 29 & 0.6 & 0.34  & 13    & 0.2   & 0.00 \\
\mbox{SPS 4} &  733 & 545 & 21 & 6.1 & 4.2    & 490 & 19 & 26 & 0.3 & 1.8   & 70    & 5.5   & 0.06 \\
\mbox{SPS 5} &  722 & 262 & 20 & 6.1 & 23     & 490 & 19 & 26 & 1.6 & 2.4   & 77    & 1.2   & 0.02 \\
\mbox{SPS 6} &  720 & 503 & 20 & 6.1 & 11     & 490 & 19 & 26 & 0.5 & 2.0   & 79    & 2.5   & 0.03 \\
\mbox{SPS 7} &  950 & 807 & 25 & 7.5 & 8.4    & 530 & 21 & 29 & 0.2 & 0.30  & 12    & 0.5   & 0.00 \\
\mbox{SPS 8} &  839 & 978 & 23 & 6.8 & 0.034  & 510 & 20 & 27 & 1.7 & 0.81  & 28    & 22    & 0.10 \\
\mbox{SPS 9} & 1182 & 930 & 30 & 8.8 & 9.1    & 570 & 23 & 31 & 0.3 & 0.027 & 2.2   & 0.1   & 0.00 \\
\hline     & 300 & 300    & 11 & 3.1 & 0      & 370 & 15 & 19 & 0   & 500   & 14000 & 13000 & 780  \\
\mbox{Toy} & 300 & \infty & 11 & 3.1 & 0      & 350 & 14 & 19 & 19  & 900   & 15000 & 15000 & 240  \\
\mbox{models} & 800 & 800 & 22 & 6.6 & 0      & 500 & 20 & 27 & 0   &  1.6  & 39    & 38    & 0.28 \\
           & 800 & \infty & 22 & 6.6 & 0      & 500 & 20 & 27 & 27  &  3.8  & 54    & 51    & 2.6  \\
\hline
\end{array}$$
\caption{Parameters of various standard MSSM points and the toy scenarios that we will analyze. The column $\l(\Gamma_\mathbf{8_A}^{\rm ann}\r)$ refers to the value without the $m_\sq$-dependent suppression factor in (\ref{8Aannihfull}). We also compare the gluinonia production cross section $\sigma^{\rm bound}$ (estimated in the narrow-width approximation) with the continuum gluino-pair production cross section $\sigma^{\rm cont}$, and the last two columns show the annihilation cross sections into dijets and $t\bar t$ (while the rest of $\sigma^{\rm bound}$ corresponds to single gluino decays like in the continuum).}
\label{tab-gogo-SPS}
\end{table}

In the narrow width approximation, the production cross sections for gluinonia are simply related to their decays through (\ref{sigma-hat-general-total}):
\bea
\hat\sigma_{gg\to \mathbf{1}}(\hat s)
&\simeq& \frac{\pi^2\,\Gamma_{\mathbf{1}\to gg}}{8M_\mathbf{1}}\, \delta\l(\hat s - M_\mathbf{1}^2\r)
\simeq \frac{243\pi^2}{64}\alpha_s^2\bar\alpha_s^3\, \delta\l(\hat s - M^2\r) \,,\nn\\
\hat\sigma_{gg\to \mathbf{8_S}}(\hat s)
&\simeq& \frac{\pi^2\,\Gamma_{\mathbf{8_S}\to gg}}{M_\mathbf{8_S}}\,\delta\l(\hat s - M_\mathbf{8_S}^2\r)
\simeq \frac{243\pi^2}{256}\alpha_s^2\bar\alpha_s^3\, \delta\l(\hat s - M^2\r) \,,\nn\\
\hat\sigma_{q_i\bar q_i\to \mathbf{8_A}}(\hat s)
&\simeq& \frac{32\pi^2\,\Gamma_{\mathbf{8_A}\to q_i\bar q_i}}{3M_\mathbf{8_S}}\,\delta\l(\hat s - M_\mathbf{8_A}^2\r)\nn\\
&\simeq& \frac{9\pi^2}{8} \sum_{\chi=1,2}\left(\frac{m_{\sq_{i,\chi}}^2 - m_\go^2}{m_{\sq_{i,\chi}}^2 + m_\go^2}\right)^2 \alpha_s^2\bar\alpha_s^3\, \delta\l(\hat s - M^2\r) \,,
\label{gluinonium-sigmahat}
\eea
where the sum in the last line is over squark chiralities.

Alternatively, the expressions for the cross section can be derived using the near-threshold tree-level gluino pair production cross sections, which are (see, e.g.,~\cite{Beenakker:1996ch})
\be
\frac{\hat\sigma_0(gg\to\go\go)}{\beta} = \frac{27\pi\alpha_s^2}{64 m_\go^2}\,, \qqqq
\frac{\hat\sigma_0(q_i\bar q_i\to\go\go)}{\beta} = \frac{\pi\alpha_s^2}{6m_\go^2}\sum_{\chi=1,2}\left(\frac{m_{\sq_{i,\chi}}^2 - m_\go^2}{m_{\sq_{i,\chi}}^2 + m_\go^2}\right)^2 \,.
\ee
In the $gg$ channel $1/6$ of the production is in the color-singlet representation, $1/3$ in the color-octet $\mathbf{8_S}$, and $1/2$ in $\mathbf{27}$ (which does not bind), while in the $q\bar q$ channel all the production is in the color-octet $\mathbf{8_A}$~\cite{Hagiwara:2009hq}. Then, using Eq.~(\ref{sigma-Phi1/Phi2}) from Appendix~\ref{sec-gen-bound},
with the wavefunctions in Eq.~(\ref{hydrogen-like})  computed with the color factors of Eq.~(\ref{gluinonium-color-factors}) appropriate for each representation, we obtain Eq.~(\ref{gluinonium-sigmahat}).

The radial excitations are described by the same expressions as (\ref{gluinonium-sigmahat}),
but with the annihilation widths scaled by
\be
\Gamma^{\rm ann} \to \frac{\Gamma^{\rm ann}}{n^3} \,.
\ee
In the limit $\Gamma_\go\to 0$ the radial excitations can be summed to effectively change $\Gamma^{\rm ann}$ by an overall prefactor
\be
\sum_{n=1}^\infty \frac{1}{n^3} = \zeta(3) \simeq 1.202 \,.
\label{radial-sum}
\ee
For a finite $\Gamma_\go$ the contribution of the radial excitations is somewhat smaller.\footnote{For finite $\Gamma_\go$, the
excited modes, with their $1/n^3$-reduced annihilation rate, still need to compete with the same $2\Gamma_\go$. The correct
correction can be easily computed numerically. One can also take into account that some of the
 excited states may decay into $P$ states whose annihilation rates are small. This reduces their
 contribution to the annihilation signal. We do not take this possibility into account. We also did not take into account the direct production of $P$ states and higher angular excitations.}

The total production cross sections are given by
\be
\sigma(pp \to \mathbf{1}) = \zeta(3)\frac{\pi^2}{8 M_\mathbf{1}^3}
\mathcal{L}_{g g}(M_\mathbf{1}^2)\,\Gamma_{\mathbf{1}\to gg} \,,
\label{hadron-cs-1}
\ee
\be
\sigma(pp \to \mathbf{8_S}) = \zeta(3)\frac{\pi^2}{M_\mathbf{8_S}^3}
\mathcal{L}_{g g}(M_\mathbf{8_S}^2)\,\Gamma_{\mathbf{8_S}\to gg} \,,
\ee
\be
\sigma(pp \to \mathbf{8_A}) = \zeta(3)\frac{64\pi^2}{3M_\mathbf{8_A}^3}
\sum_q
\mathcal{L}_{q \bar{q}}(M_\mathbf{8_A}^2)\,
\Gamma_{\mathbf{8_A}\to q\bar q} \,,
\label{cs-8A}
\ee
where the luminosities are
\be
\mathcal{L}_{i j}(\hat s) = \frac{\hat s}{s}\,\int_{\hat s/s}^1 \frac{\rd x}{x}\, f_i(x)\, f_j\l(\frac{\hat s}{xs}\r) \,, \label{lumi}
\ee
where $i$ and $j$ denote the initial state partons and $\sqrt{s}=E_\mathrm{CM}$ is the collider energy.

\begin{figure}[t]
\begin{center}
\includegraphics[width=0.7\textwidth]{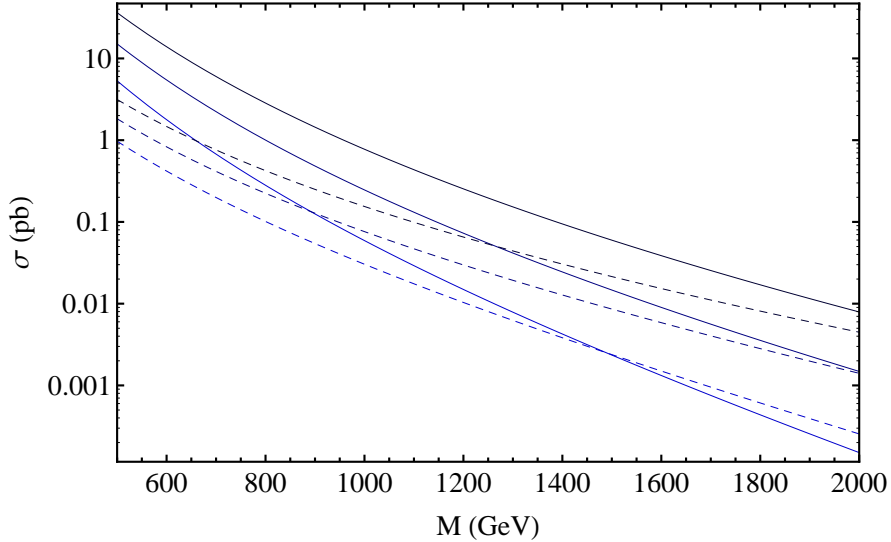}
\caption{Gluinonia production cross section at the LHC as a function of the gluinonium mass $M \simeq 2m_\go$. Solid lines refer to the sum of $\mathbf{1}$ and $\mathbf{8_S}$ pseudoscalar gluinonia which annihilate into $gg$ and $t\bar t$, and dashed lines refer to $\mathbf{8_A}$ vector gluinonium which annihilates into $q\bar q$ (including $t\bar t$). The three lines in each case correspond,
from top to bottom, to LHC center-of-mass energy of $\sqrt{s} = 14$, $10$ and $7$ TeV.}
\label{fig-cs}
\end{center}
\end{figure}

These cross sections are shown in Figure~\ref{fig-cs} as a function of the gluinonium mass for the LHC at $\sqrt{s}=$ 14 TeV, 10 TeV
and 7 TeV. For this plot, we have used the LO MSTW 2008 parton distribution functions~\cite{Martin:2009iq} evaluated at the scale
$M = 2 m_\go$ and neglected both the gluino width $\Gamma_\go$ and the suppression of $\Gamma_{\mathbf{8_A}\to q\bar q}$ due to the finite squark masses.
These cross sections sum to a few percent of the total gluino pair production cross section, fairly independently of the gluino mass over this range. We will discuss this further in Section~\ref{sec-annih-rate}.

\begin{table}[t]
$$\begin{array}{|c|c|c|c|c|c||c|c|}\hline
\,    m_\go
\,&\, m_\sq
\,&\, \mathbf{1}, \mathbf{8_S} \to gg
\,&\, \mathbf{1}, \mathbf{8_S} \to t\bar t
\,&\, \mathbf{8_A} \to q\bar q
\,&\, \mathbf{8_A} \to t\bar t
\,&\, \mbox{any} \to gg,q\bar q
\,&\, \mbox{any} \to t\bar t\,\\
   \, \mbox{GeV}
\,&\, \mbox{GeV}
\,&\, \mbox{fb}
\,&\, \mbox{fb}
\,&\, \mbox{fb}
\,&\, \mbox{fb}
\,&\, \mbox{fb}
\,&\, \mbox{fb} \\
\hline
300 & 300    & 13000 & 780  & 0    & 0   & 13000 & 780  \\\hline
300 & \infty & 14000 & 0    & 1200 & 240 & 15000 & 240  \\\hline
800 & 800    & 38    & 0.28 & 0    & 0   &    38 & 0.28 \\\hline
800 & \infty & 38    & 0    & 13   & 2.6 &    51 & 2.6  \\\hline
\end{array}$$
\caption{Gluinonium annihilation cross sections at the LHC.}
\label{tab-cs}
\end{table}

The cross sections for sample masses $m_\go = 300$ and $800$ GeV, with squarks either degenerate with the gluinos or much heavier, are shown in Table~\ref{tab-cs}.
These numbers differ somewhat from other results in the literature.
Ref.~\cite{Cheung:2004ad} has found cross sections which
are a factor of $2-3$ smaller than our results.
This happens to a large extent because they use $\alpha_s^3$ instead of $\bar\alpha_s^3$ in their expressions for $|\psi(\vv{0})|^2$. The results of~\cite{Chikovani:1996bk} for the pseudoscalar gluinonium are about $4$ times bigger than ours (even though they applied certain cuts while we did not). Because of the factor-of-$8$ error in the expression they use for $\Gamma_{\mathbf{8_S}\to gg}$ as we explained in Footnote~\ref{error-explanation}, their combined cross section for $\mathbf{1}$ and $\mathbf{8_S}$ is indeed expected to be by a factor of about $2.4$ larger. However, it also seems that they used $\alpha_s$ instead of $\bar\alpha_s$, which should have made their result smaller. But they have included a $K$-factor
of $2.0$ which can compensate for that. A $K$-factor does need to be included in principle, but its size is still unknown. The effect of initial-state radiation on the gluinonium production has been computed in~\cite{Hagiwara:2009hq}, giving a near-threshold $K$-factor that varies from $1.2$ for $m_\go = 200$ GeV to $1.8$ for $m_\go = 1$ TeV. However, this does not yet include the emission of non-collinear gluons which can give an additional correction of the same order~\cite{Hagiwara:2009hq}. Other corrections which can modify the results by $\sim 50\%$ are higher-order QCD corrections to the potential (\ref{V(r)}), a confining term that may still affect the result for the smaller masses, and relativistic corrections suppressed by $v^2$. Computations of part of the QCD corrections have been recently presented in~\cite{Kauth:2009ud}.
Furthermore, we need to include corrections also to the annihilation process (final-state radiation etc.). Since some of these quantities have not been computed yet, we will not include a $K$-factor in our simulation of the gluinonium signal. We will not include a $K$-factor for the QCD background either.

\section{Gluinonium: simulation\label{sec-gogo-sim}}
In the previous section, we found that 300 GeV gluinos can have
a cross section for annihilation decay as high as 15 pb. Although the cross sections shrink sharply with gluino mass, it is certainly possible that the annihilation decays of gluinonia might be seen.
The dominant decay mode of the gluinonia is to dijets, however
the decays to $t\bar{t}$ may be easier to see because of smaller backgrounds. In order
to determine what we might learn about the underlying physics if these resonances can be found, we
now examine the dijet and $t\bar{t}$ signals and backgrounds through Monte Carlo simulation.

We simulate both the signals and backgrounds at the LHC at 14 TeV
using \textsc{Pythia} version 8.120~\cite{Sjostrand:2006za,Sjostrand:2007gs}
with MSTW2008 LO parton distribution functions~\cite{Martin:2009iq}. For clustering the particles into jets, we use the \textsc{SISCone}
jet algorithm version 2.0.1~\cite{Salam:2007xv} with radius $R = 1$ for the dijet analyzes and $R=0.5$ for the $t\bar t$
analysis.

\subsection{Signal and background\label{sec-simulation}}
As we saw in the previous section, the bound states of different spin have different production and decay modes.
The pseudoscalars, which include a color-singlet, $\Phi_\mathbf{1}$, and a color-octet, $\Phi_\mathbf{8_S}$,
couple to both gluons and heavy quarks.
In models for which the annihilation decays are relevant, the resonances have very small widths. Then we
can model them with effective interactions of the form
\begin{align}
{\cal L}_{\mathbf{1}} &=
c_{gg}^\mathbf{1}\frac{1}{m_\go} \epsilon^{\mu\nu\rho\sigma}\Phi_\mathbf{1} G_{\mu\nu}^a G_{\rho\sigma}^a+
c_{q\bar{q}}^\mathbf{1} \frac{m_q}{m_\go}\,\Phi_\mathbf{1}\,i\bar q\gamma^5 q \,,\\
{\cal L}_\mathbf{8_S} &=
c_{gg}^\mathbf{8_S}\frac{1}{m_\go}\, \epsilon^{\mu\nu\rho\sigma}d^{abc}\,\Phi_\mathbf{8_S}^a G_{\mu\nu}^b G_{\rho\sigma}^c +
c_{q\bar{q}}^\mathbf{8_S}
\frac{m_q}{m_\go}\,\Phi_\mathbf{8_S}^a \l(T^a\r)_i^j\,i\bar q^i \gamma^5 q_j \,,
\end{align}
where $a,b,c$ and $i,j$ are color indices of the adjoint and fundamental representations, respectively.
The factor of $m_q$ in the normalization of the $q\bar{q}$ operators is chosen so that they vanish as $m_q\to0$, as they must by chirality (the rates are approximately proportional to $m_q^2$: see (\ref{1annih-ttbar}) and (\ref{8Sannih-ttbar})).
The vector, $V^\mu_\mathbf{8_A}$, has suppressed couplings to gluons and its coupling to quarks can be described by the effective interaction
\be
{\mathcal L}_\mathbf{8_A} =
c_{q\bar{q}}^\mathbf{8_A}
\l(V^\mu_\mathbf{8_A}\r)^a \l(T^a\r)_i^j\,\bar q^i \gamma_\mu q_j \,.
\ee
 All of the coefficients
$c_{gg}^\mathbf{R}$ and $c_{q\bar{q}}^\mathbf{R}$ are fixed by matching to the annihilation rates.
One could also write down operators with additional derivatives,
but since in the narrow width approximation the momenta of the incoming particles are fixed, we can simply absorb the effect of these additional operators into the normalization.

As a simplifying assumption, since the color matrices do not affect the geometry of the processes, at least at leading order, we treat all the gluinonia as color-singlets. This facilitates the Monte Carlo simulation.
In particular, the pseudoscalar gluinonia reduce to a pseudoscalar Higgs
with appropriately modified couplings. The vector gluinonium can be simulated as a vector $Z'$ with flavor universal couplings of appropriate strength (to the vector current of quarks only).
We restrict our discussion of the simulation results to the representative cases described in Table~\ref{tab-cs}, with $m_\go=$ 300 GeV ($M=600$ GeV) and $m_\go=800$~GeV ($M=1600$~GeV).

The background for the dijet signals ($gg$ and $q\bar q$) is QCD dijets.
For the $t\bar t$ channel background, we use the standard model $t\bar t$ production, including both
$gg\to t\bar t$ ($\sim 85\%$) and $q\bar q \to t\bar t$ ($\sim 15\%$) and also single top production (predominantly $qq \to tq$).
 We do not include fakes from QCD or $W/Z+\mbox{jets}$ since it seems likely the experiments
 can distinguish these backgrounds from $t\bar{t}$ with good efficiency~\cite{CMS-PAS-TOP-09-009}, but in any case, top-tagging is a subject largely orthogonal to the current work.

\subsection{Analysis of the dijet channel\label{sec-dijet}}

\begin{figure}[t]
\begin{center}
\psfrag{A}[][][0.8]{100 pb${}^{-1}$}
\psfrag{B}[][][0.8]{1 fb${}^{-1}$}
\psfrag{C}[][][0.8]{10 fb${}^{-1}$}
\psfrag{D}[][][0.8]{$pp \to (\go \go) \to \mathrm{dijets}$}
\psfrag{E}[][][0.8]{$pp \to (\go \go)_{\rm new} \to \mathrm{dijets}$}
\includegraphics[width=0.7\textwidth]{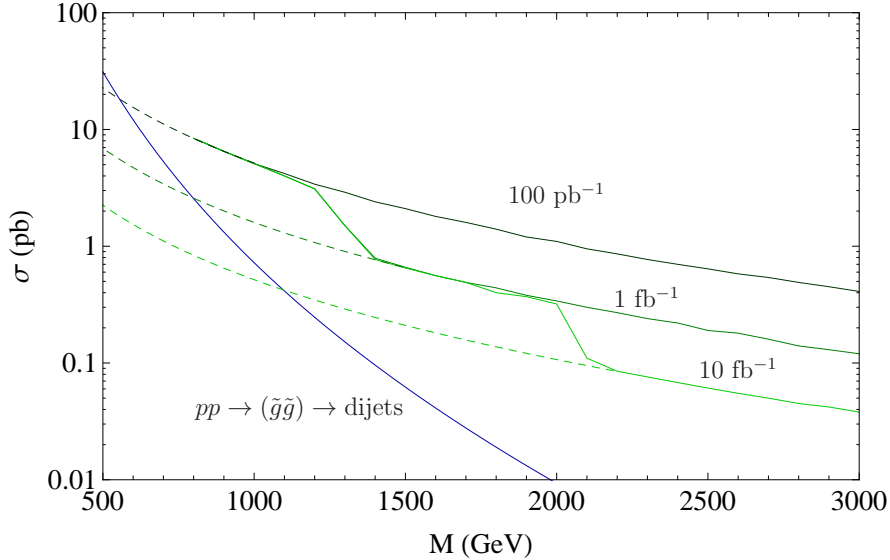}
\end{center}
\caption{Dijet search reach ($95\%$ confidence level exclusion) at {\sc CMS}~\cite{CMS-NOTE-2006/070} and the gluinonium signal with $|\eta_{\rm jet}| < 1$. The steps are due to prescaling.
The dashed lines are our extrapolations assuming that the prescaling can be reduced by adjusting the triggers.}
\label{fig-cmsreach}
\end{figure}

First, let us consider the dijet signatures of annihilation decays. These come from
 pseudoscalars decaying to $gg$ and vectors decaying to $q\bar{q}$.
For the $m_\go = 300$ GeV case, the jets will dominantly have $p_T \gtrsim 200$~GeV,
and for the $m_\go=800$~GeV case $p_T \gtrsim 500$~GeV.
For the light gluino, the $200$ GeV $p_T$ cut is problematic because it will not survive the high-luminosity triggers of either {\sc ATLAS} or {\sc CMS}. For example, with an instantaneous luminosity of $10^{34}$ cm${}^{-2}$ s${}^{-1}$,
only dijet events with $p_T> 450 $ GeV can survive prescaling.
As pointed out in~\cite{CMS-NOTE-2006/070}, if we already have information about the gluino mass, for example from cascade decays,
than the high-level trigger can be adjusted to reduce the prescaling by a factor of up to 1000.
Therefore, it is still possible that the dijet signatures of the 600 GeV gluinonium annihilation decays may be
found. In Figure~\ref{fig-cmsreach}, we compare the  dijet reach at {\sc CMS} (for instantaneous luminosities of $10^{32}$, $10^{33}$ and $10^{34}$ cm${}^{-2}$ s${}^{-1}$) to the gluinonium cross section. The sawtooth
shape of the reach is due to prescaling, so we have extrapolated these curves to lower mass, assuming
the prescaling can be overcome by adjusting the high-level trigger.

\begin{figure}[t]
\begin{center}
\includegraphics[scale=0.41]{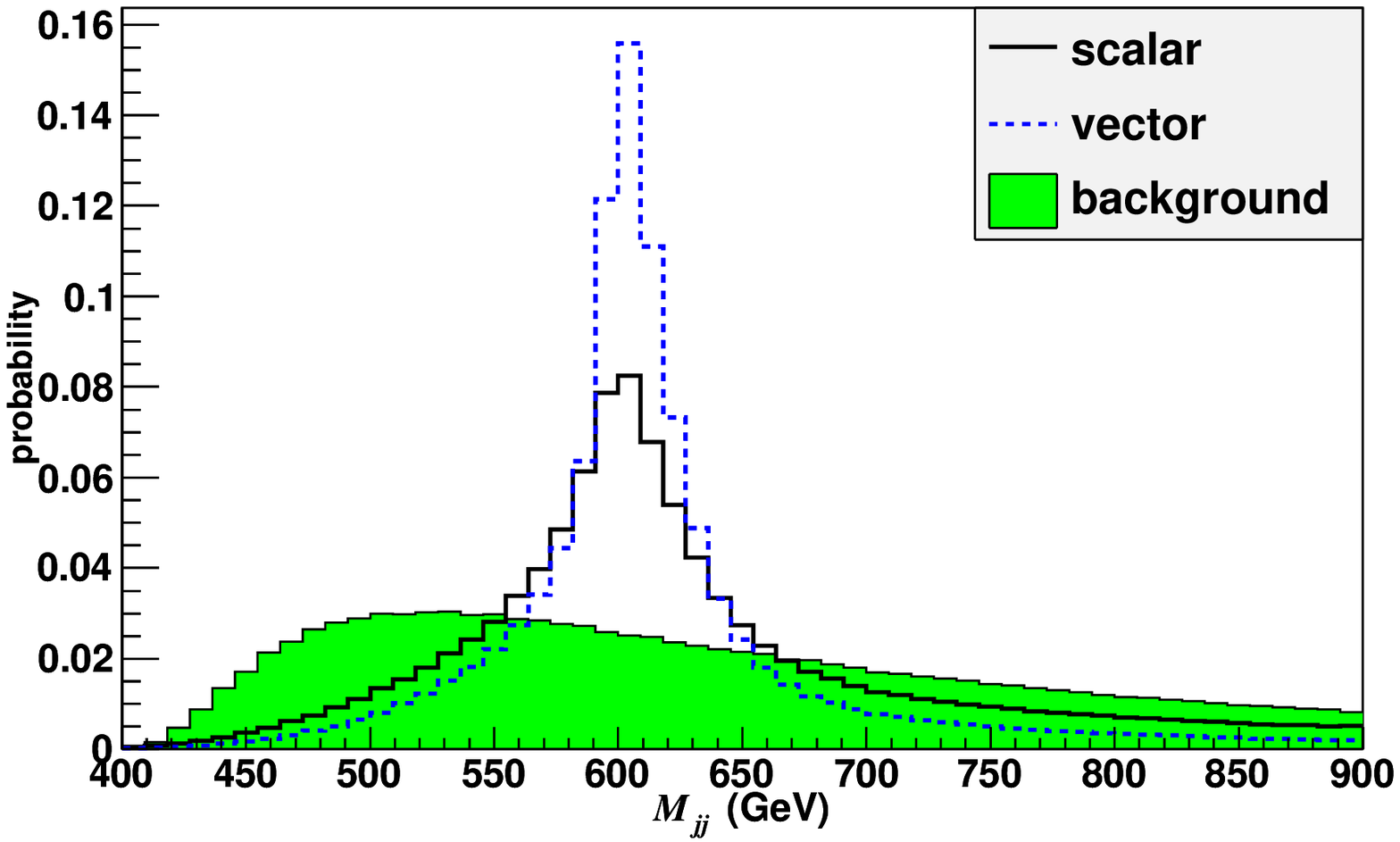}
\includegraphics[scale=0.41]{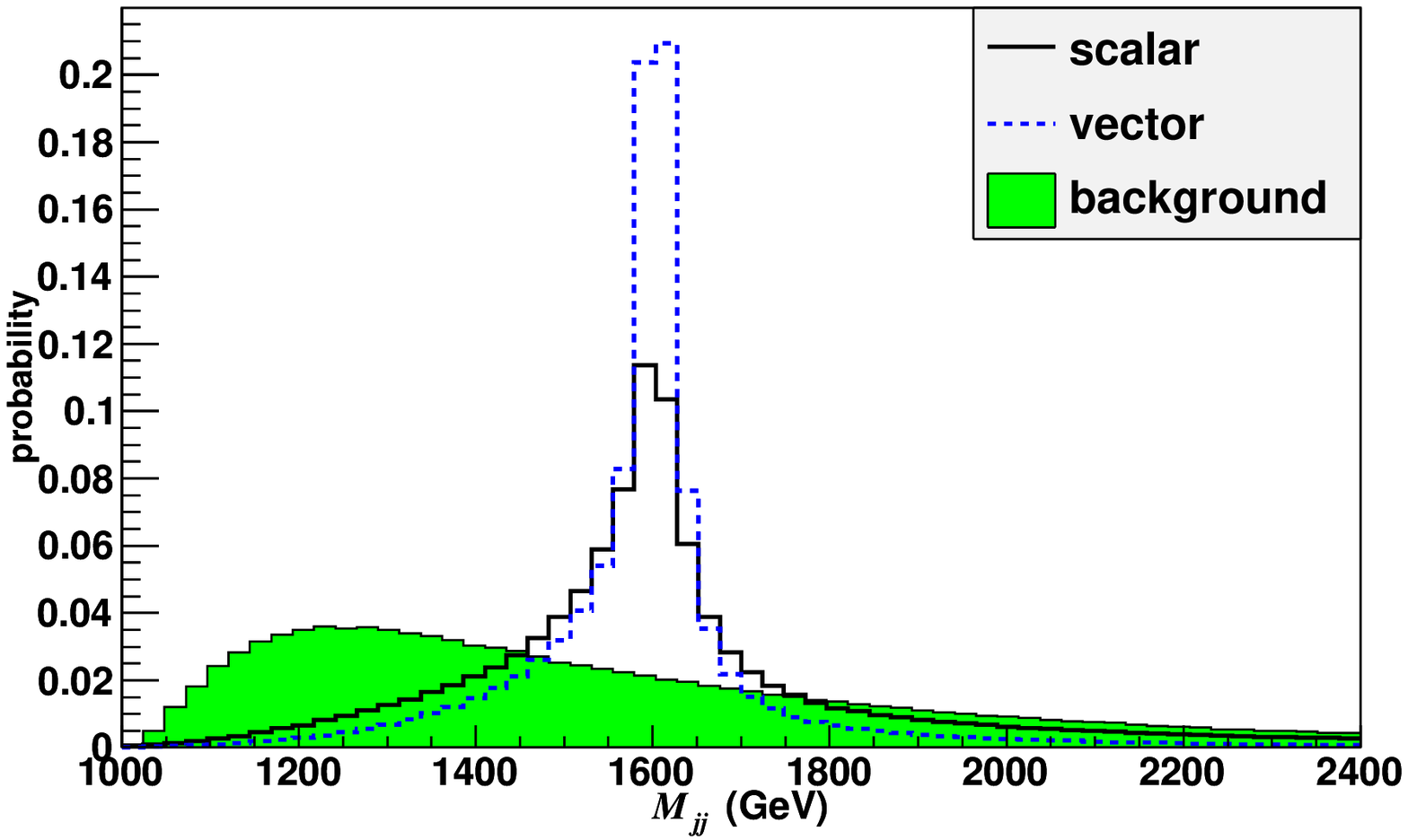}
\end{center}
\caption{The invariant mass of the two hardest jets. Left is the $300$ GeV gluino and background
after a $p_T^{\mathrm{jet}}>200$ GeV cut, and right is the $800$ GeV gluino and background with a $p_T^\mathrm{jet}>500$ GeV cut.}
\label{fig-mass}
\end{figure}

To study the distributions, we simulate the signals and the backgrounds requiring two jets with $p_T>200$ GeV
for the $m_\go=300$ GeV case and two jets with $p_T>500$ GeV for the $m_\go=800$ GeV case.
Figure~\ref{fig-mass} shows the distribution of the invariant mass of the two hardest jets in these samples.
Note that the signal, whose intrinsic width is a fraction of a GeV, broadens to about $5\%$ of the mass of the resonance due to QCD radiation (the scaling of the radiation width with the mass of the resonance follows by dimensional analysis, since QCD is almost conformal at these energies).
The enormous background and trigger issues (for the light gluino case) notwithstanding, the gluino mass can, in principle,
be determined from the location of dijet invariant mass peak.

\begin{figure}[t]
\begin{center}
\includegraphics[scale=0.41]{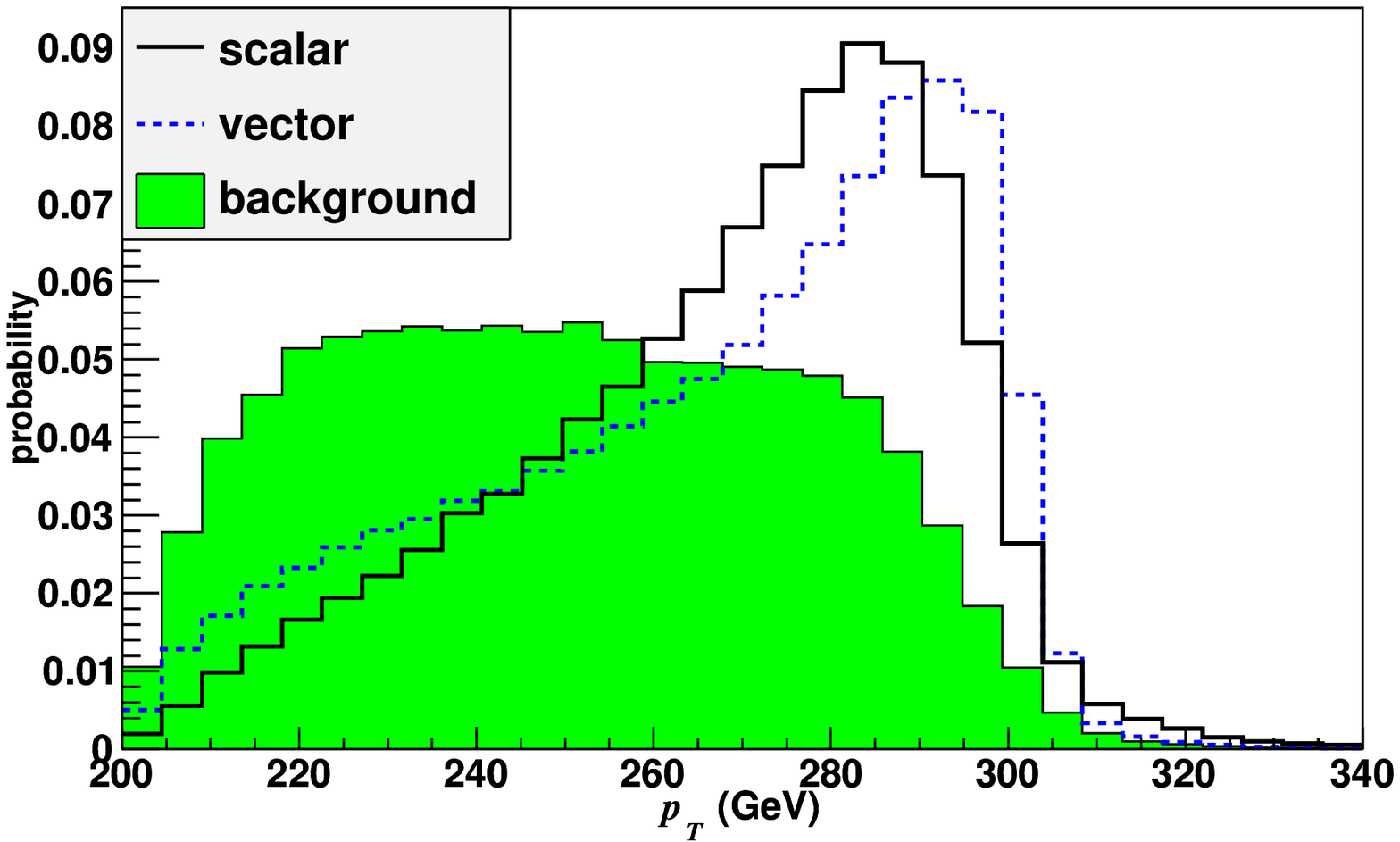}
\includegraphics[scale=0.41]{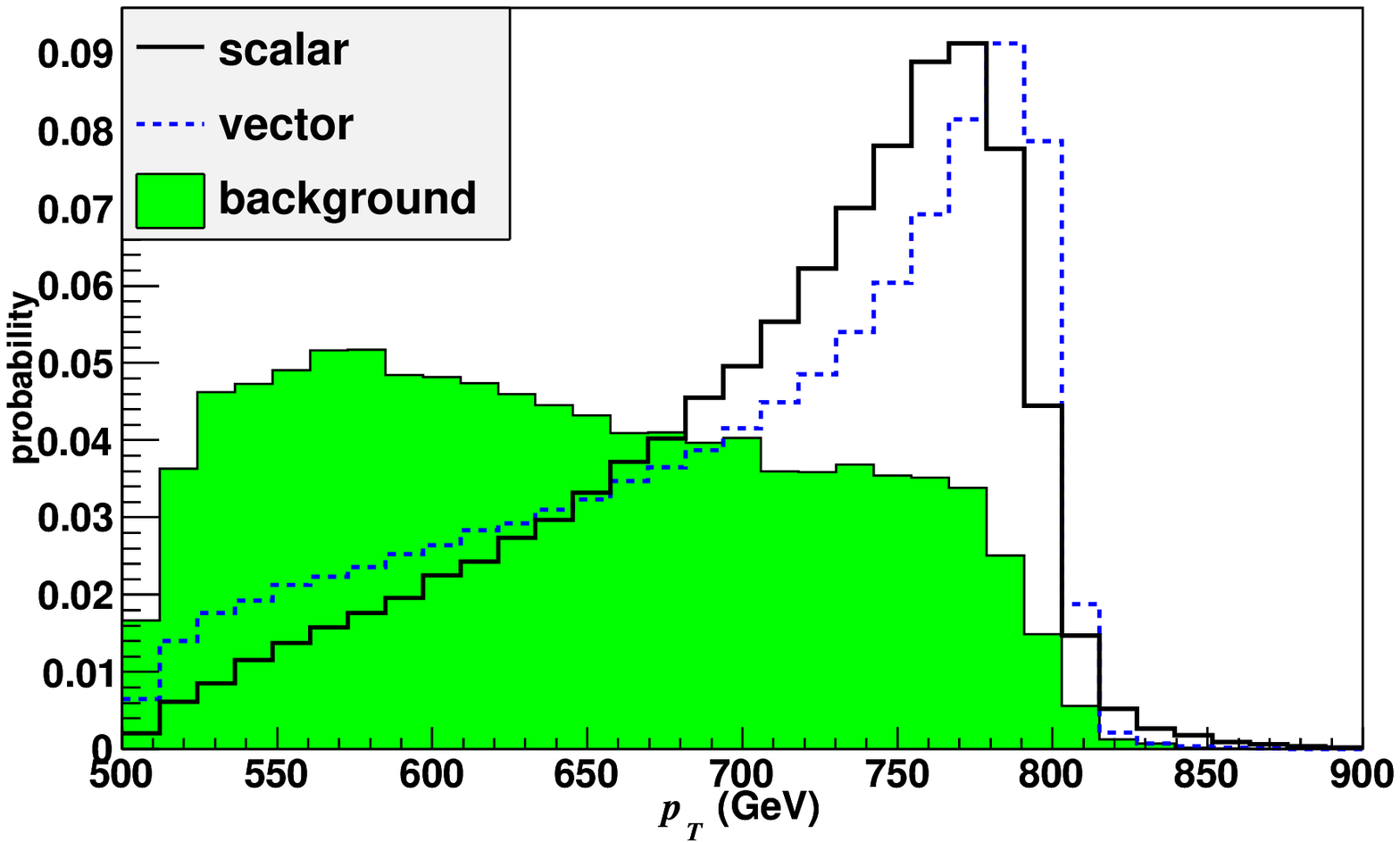}\\
\includegraphics[scale=0.41]{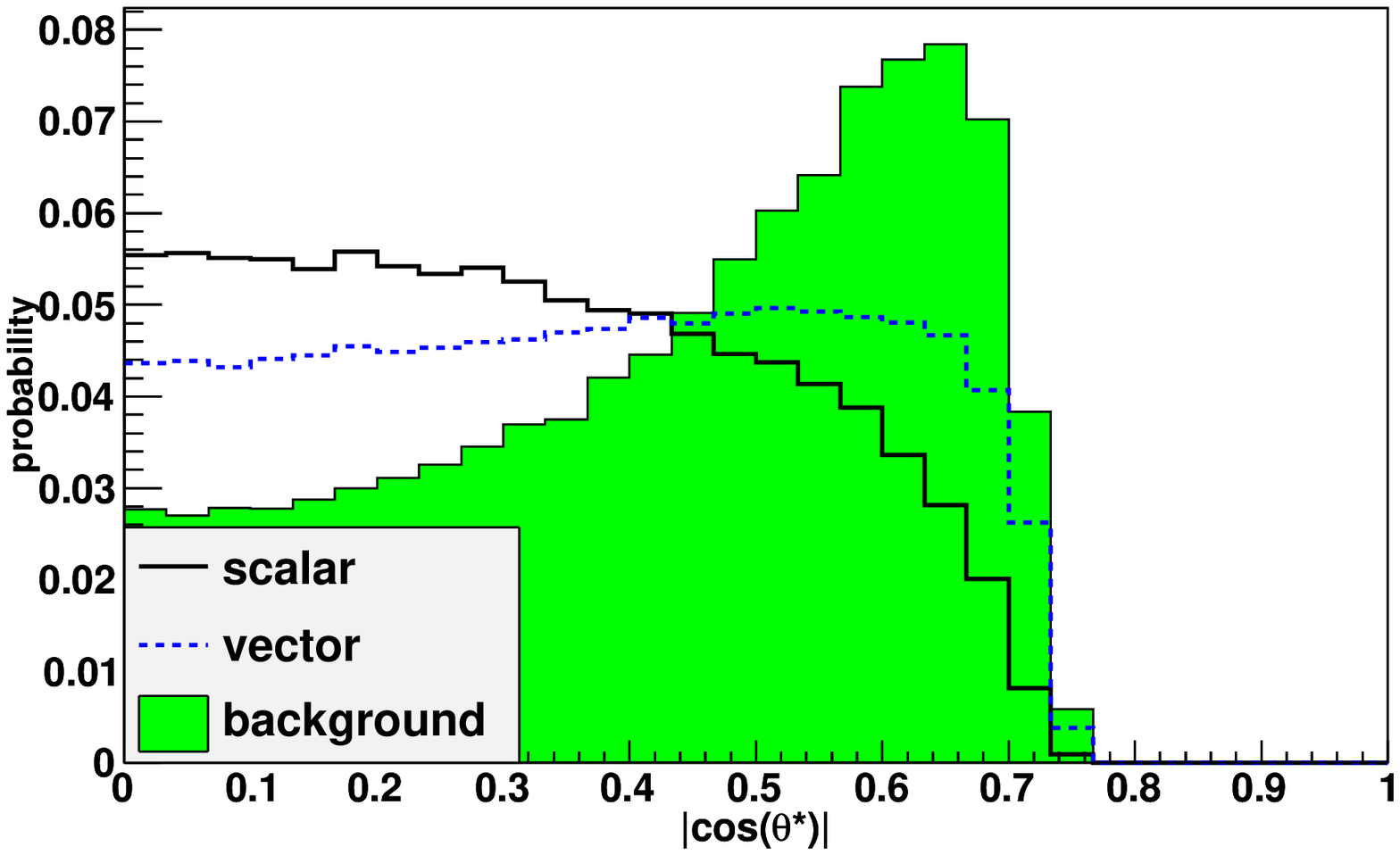}
\includegraphics[scale=0.41]{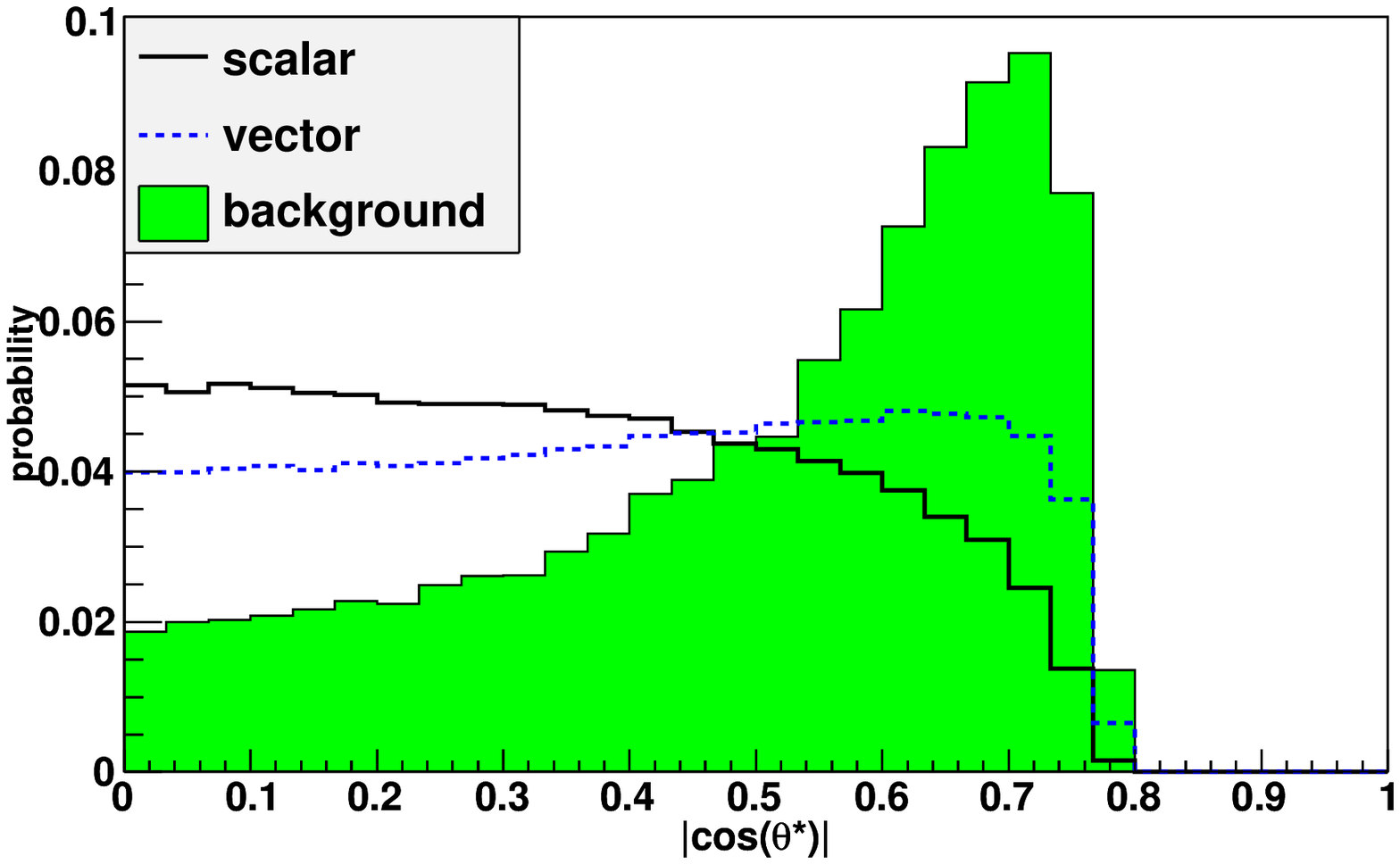}
\caption{Distributions of the average transverse momentum of the two hardest jets (top row) and the scattering angle in the collision frame
(bottom row), in invariant mass range $580\mbox{ GeV} < M_{jj} < 620\mbox{ GeV}$ for the $300$ GeV gluino case (left) and in invariant mass range $1560\mbox{ GeV} < M_{jj} < 1640\mbox{ GeV}$ for the $800$ GeV gluino (right).}
\label{fig-pT-angle}
\end{center}
\end{figure}

Let us now study more closely the behavior of the signal and the background near the peak, in the window of $580\mbox{ GeV} < M_{jj} < 620\mbox{ GeV}$ for $m_\go = 300$ GeV and
$1560\mbox{ GeV} < M_{jj} < 1640\mbox{ GeV}$ for  $m_\go=800$ GeV.
In Figure~\ref{fig-pT-angle}, we show the distributions of $p_T$ and the scattering angle $\cos\theta^\star$.
As expected and can be seen from these plots, the angular distribution is sensitive to the spin of the resonance.
In principle, this information can be used to verify that the resonance is indeed the gluinonium.
The experiment will measure the sum of contributions from two different spins, but as can be seen from Table~\ref{tab-cs} or Figure~\ref{fig-cs},
the spin-$0$ contribution will typically dominate.

To explore the signal and background rates more quantitatively, consider the $300$ GeV gluino. In the dijet invariant mass window $580\mbox{ GeV} < M_{jj} < 620\mbox{ GeV}$, including both the spin-0 and spin-1 resonance contributions, the signal-to-background ratio is $S/B\simeq 1\times 10^{-3}$ after the cuts $p_T > 275$ GeV, $\l|\cos\theta^\star\r| < 0.5$ for both cases from Table~\ref{tab-cs}.
This indicates that the signal extraction will be very sensitive
to theoretical and experimental systematic uncertainties of the background.
On the other hand, the statistical uncertainty on the background, characterized by $S/\sqrt{B}$, quickly shrinks
because we are dealing with a large number of events. For example, with
$\sim 10$ fb$^{-1}$ there is already a $3\sigma$ significance. Although the {\sc CMS} analysis~\cite{CMS-NOTE-2006/070} indicates that these
signals may be seen (see Figure~\ref{fig-cmsreach}),\footnote{Note however that while the analysis in~\cite{CMS-NOTE-2006/070} has taken into account the statistical noise of the background, the resolution limitations, and the various systematic uncertainties, it has not simulated the subtraction of the background from the data, which we guess is likely to be a critical issue with the small $S/B$ that we have.} enhancing the signal-to-background ratio may be critical to getting
a clean sample from which the gluinonium spin might be extracted.
Possibilities for enhancing the signal might involve jet substructure~\cite{Salam:2009jx},
color information, additional angular variables,
or multivariate techniques. These ideas are worth pursuing, but go beyond the scope of the current paper.
No matter what, detecting and characterizing the dijet annihilation decay signal of gluinonium will be a formidable challenge.

\subsection{Analysis of the $t\bar t$ channel\label{sec-ttbar}}
Another possible decay mode of gluinonium is to $t \bar{t}$. As we have shown, this decay rate depends on the masses
of the squarks as well as the gluino mass. For a 300 GeV gluino and 300 GeV squarks, the cross section times
branching ratio to $t\bar{t}$ is 780 fb. If the squarks are decoupled, the cross section is reduced to 240 fb. Although these
cross sections are small, the systematic uncertainty on the $t\bar{t}$ background is much smaller than for dijets, so
there is hope that with enough integrated luminosity, such resonances may be seen. There has also been a fair amount
of recent progress in improving the techniques to find boosted tops~\cite{Kaplan:2008ie,Giurgiu:2009wv,Thaler:2008ju,Almeida:2008tp},
which would come from very heavy gluinonium decays.

The reconstruction of the invariant mass and the other properties of the decaying resonance in the $t\bar t$ channel is more complicated than in the dijet case because the tops decay. As a result, the energy and momentum of the resonances are no longer concentrated in two hard jets, but in a combination of several jets and possibly leptons and missing energy as well.
The cleanest channel for $t\bar{t}$ is the semi-leptonic one, in which one of the tops decays hadronically and the other decays to a $b$-quark, an electron or muon, and a neutrino. The presence of the lepton (unlike in the all-hadronic channel) is very useful for rejecting the QCD background, while the fact that there is only a single neutrino (unlike in the all-leptonic channel) allows full reconstruction using the on-shell intermediate $W$. The branching ratio for semi-leptonic $t\bar{t}$ is $30\%$, and, for simplicity, we consider only this channel. The top-tagging algorithm  we use is described in Appendix~\ref{sec-ttbar-algorithm}.

\begin{figure}[t]
\begin{center}
\includegraphics[scale=0.41]{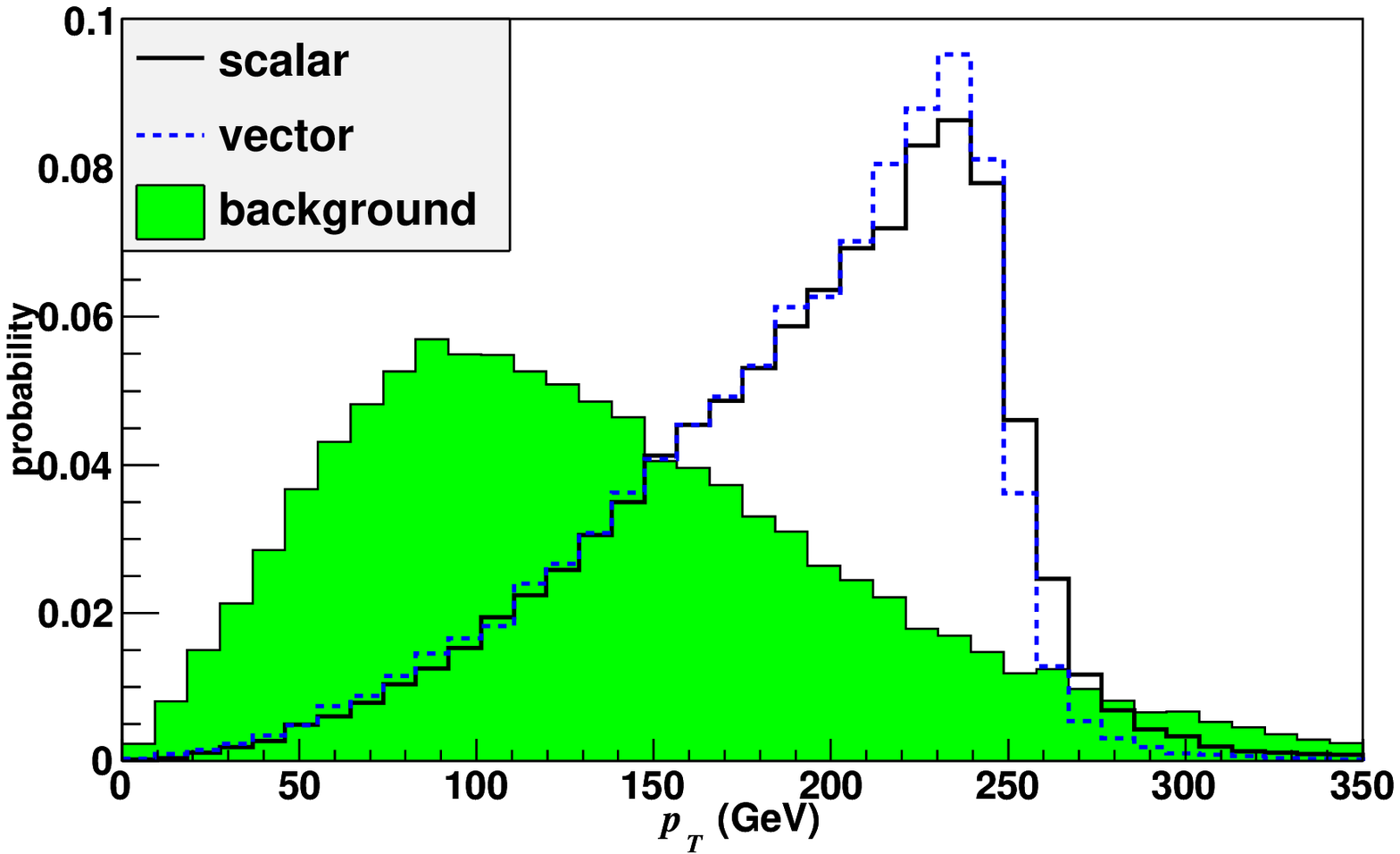}
\includegraphics[scale=0.41]{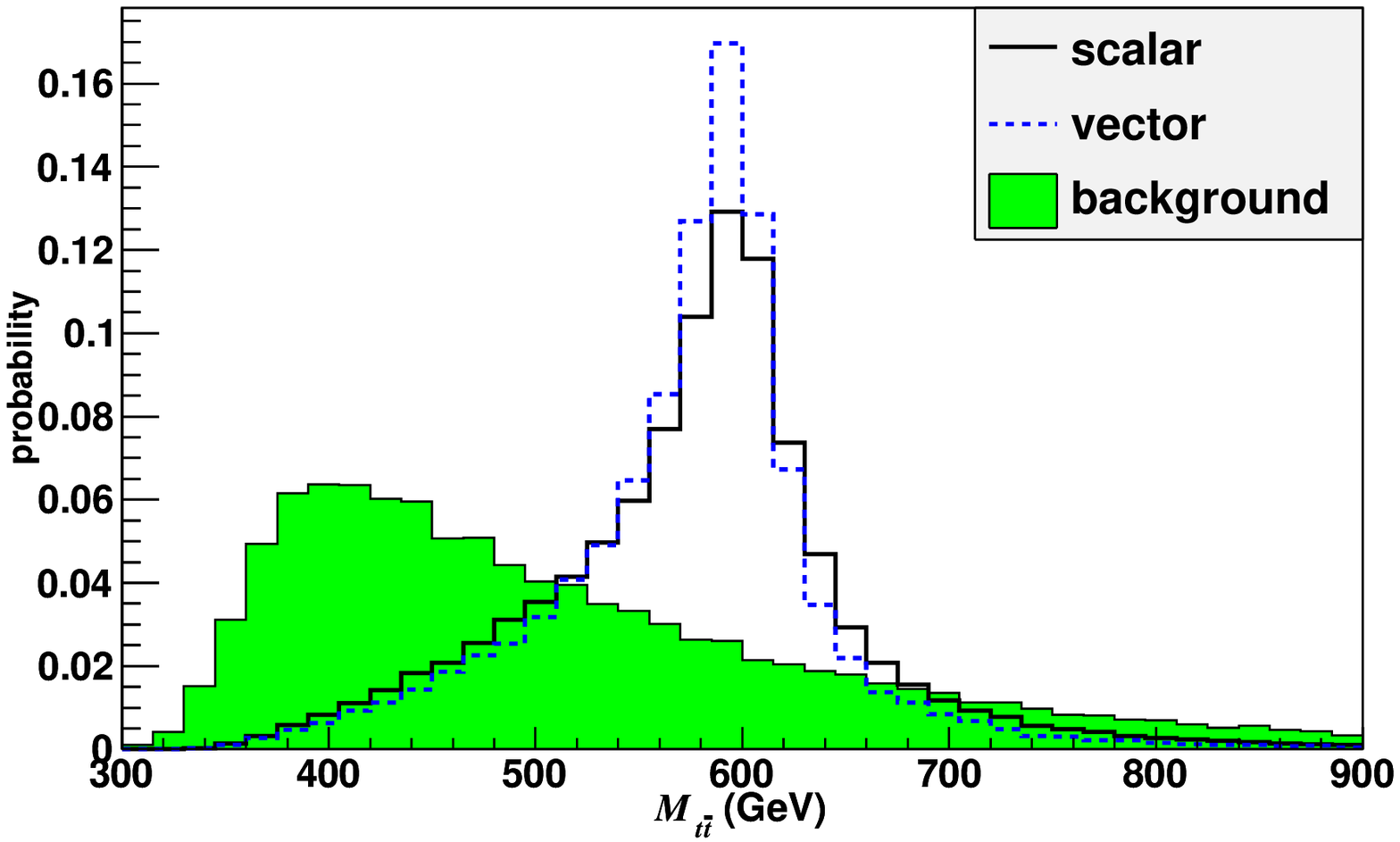}
\caption{Distributions of the average $p_T$ (left) and invariant mass (right) of the $t\bar t$ pair for a $300$ GeV gluino.}
\label{fig-ttbar-mass-pT}
\end{center}
\end{figure}

\begin{figure}[t]
\begin{center}
\includegraphics[scale=0.41]{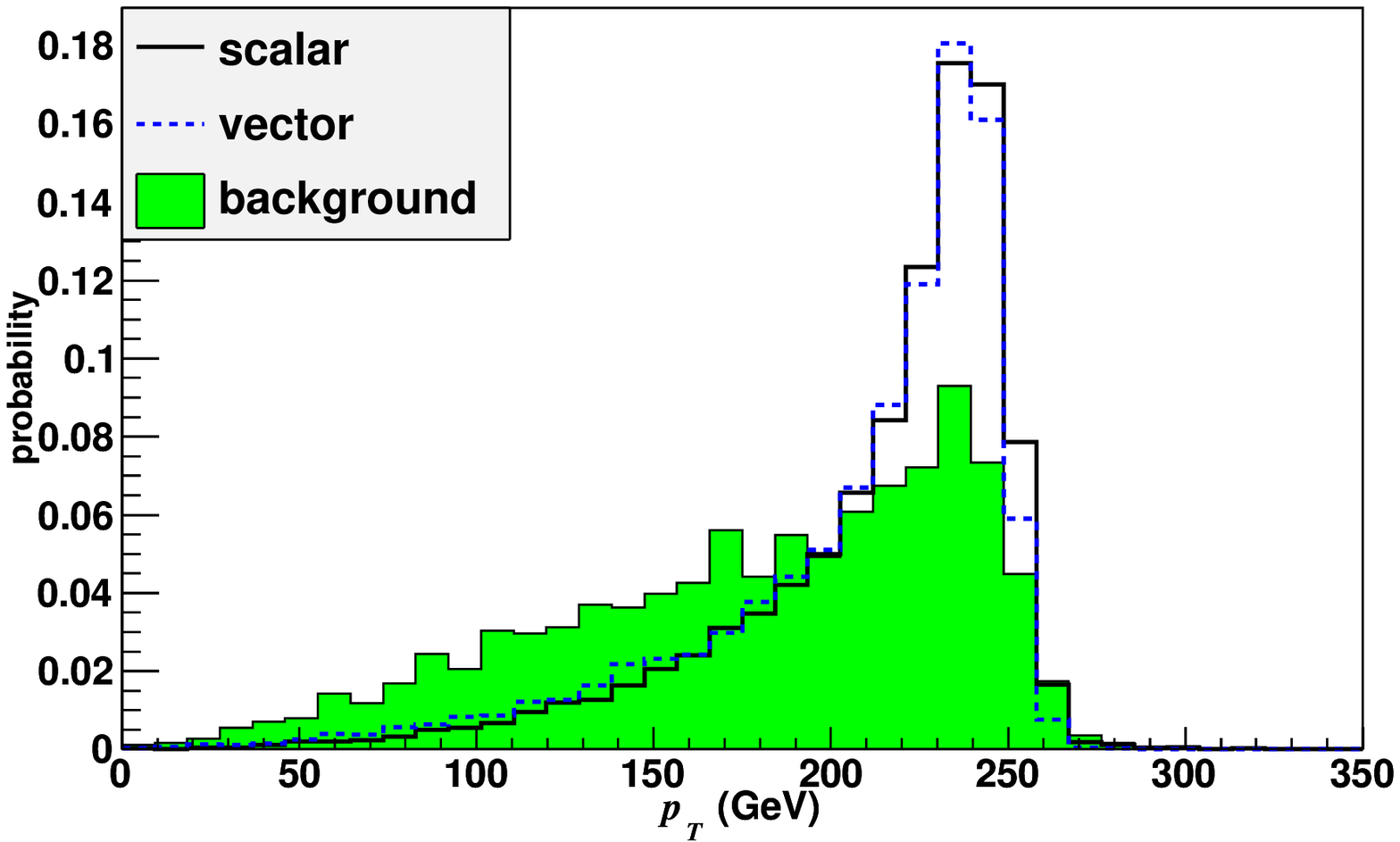}
\includegraphics[scale=0.41]{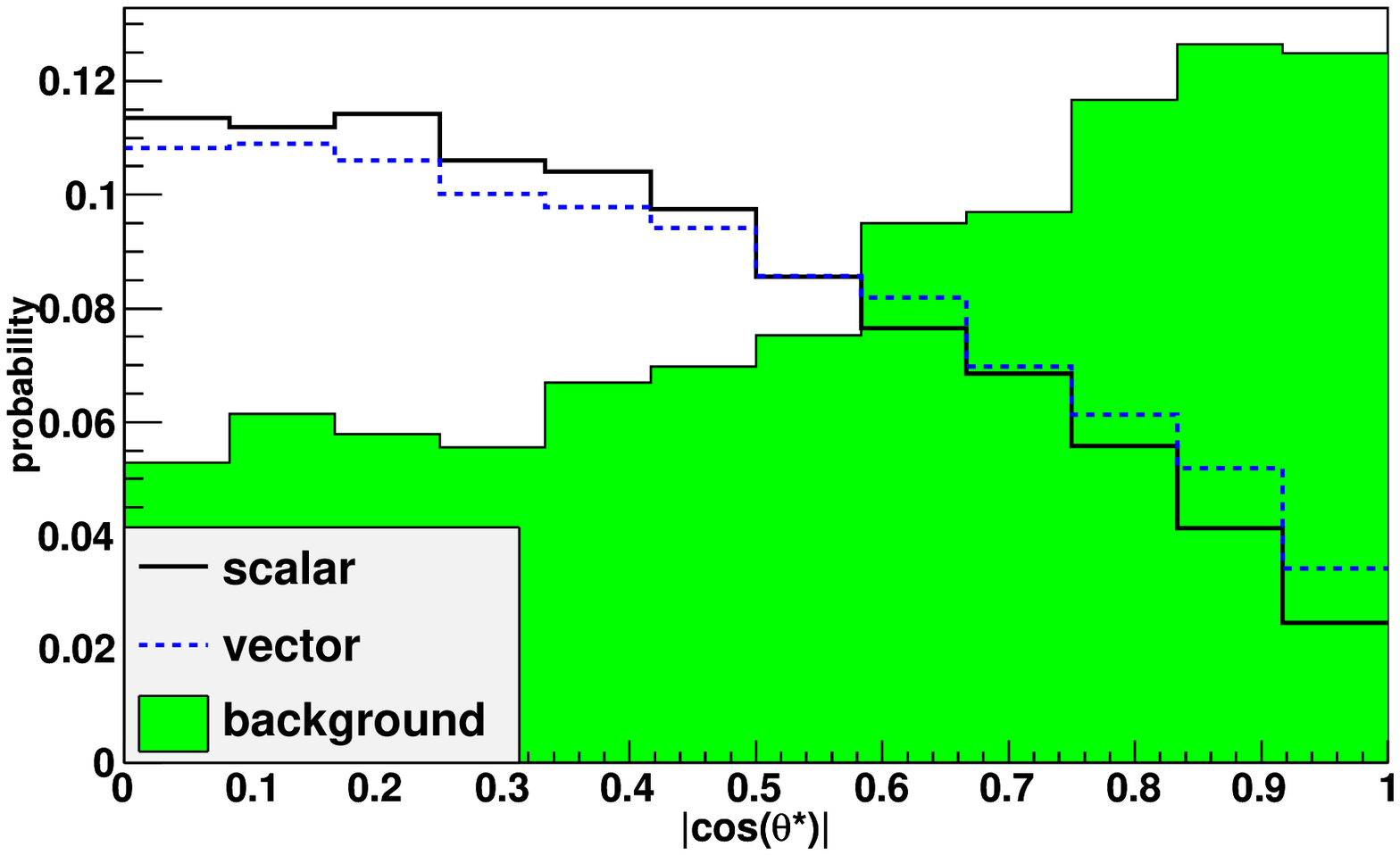}
\caption{Distributions of the average $p_T$ (left) and the scattering angle of the $t\bar t$ pair in the collision frame (right) for the $300$ GeV gluino in invariant mass range $575\mbox{ GeV} < M_{t\bar t} < 625\mbox{ GeV}$.}
\label{fig-ttbar-pT-angle}
\end{center}
\end{figure}

The resulting $t\bar{t}$ invariant mass peak is fairly sharp, as shown in Figure~\ref{fig-ttbar-mass-pT}. To study the signal in more detail
we now restrict to the mass window $575\mbox{ GeV} < M_{t\bar t} < 625\mbox{ GeV}$. The $p_T$ and $\cos\theta^\star$ distributions
in this window are shown in Figure~\ref{fig-ttbar-pT-angle}. The difference between the angular distributions of the scalar and the vector is much smaller here than in the dijet case because the top is not much lighter than the $300$ GeV gluino. In the limit of very heavy gluinonium, the differences between the spins reappear, but the cross section for
the $t\bar{t}$ signal drops too fast to make this channel useful (from Table~\ref{tab-cs}, for $m_\go = 800$ GeV, $\sigma_{t\bar t}\sim 3$ fb if the squarks are decoupled and an order of magnitude smaller if they are degenerate with the gluinos).

For the case $m_\go = m_\sq = 300$ GeV we obtain $S/B = 1.6 \times 10^{-2}$ and $S/\sqrt{B}=3$ at $500$ fb$^{-1}$ after applying the cuts $p_T > 200\mbox{ GeV}$, $\l|\cos\theta^\star\r| < 0.5$. These estimates are somewhat less promising than what has been found in
generic analysis of resonances decaying into $t\bar t$ by {\sc ATLAS}~\cite{ATL-PHYS-PUB-2009-044}.
For example, they find that $300$ fb$^{-1}$ of luminosity will allow a $5\sigma$ discovery of a $600$ GeV resonance
if its cross section is above $\sim 700$ fb, which is approximately the cross section we have.
In contrast to the dijet channel, the $t\bar{t}$ search is entirely statistics  limited. The reach could
be improved somewhat by inclusion of other $t\bar{t}$ decay channels, and through an optimization of the cuts and top-tagging algorithm.
We conclude that $t\bar{t}$ is a viable, but high luminosity, channel in which gluinonium, if it exists, could almost certainly eventually
be found.

\subsection{Estimate for the diphoton channel}

Since the gluinos are not charged under the electroweak group, the gluinonia can decay into electroweak bosons such as $\gamma\gamma$ only through loops, with branching fractions of $\sim 10^{-5}$~\cite{Kauth:2009ud}. Nevertheless, it might still be useful to consider the diphoton channel because of its much smaller background compared to the dijet and $t\bar t$ channels. We have not performed a complete analysis, but it is easy to get a rough idea by comparing to the analysis of the $\gamma\gamma$ signal of the squarkonium in~\cite{Martin:2008sv} which we will mention in the next section. It turns out that the small $\gamma\gamma$ branching ratio for the gluinonium compared to the squarkonium is roughly compensated by the larger color factors in the gluinonium production process, and the signal may be observable if the gluino is relatively light.\footnote{We thank Cliff Burgess for suggesting us to check this possibility.} For $m_\go \sim 300$ GeV, we will have $S/B$ of about $10\%$, which is much better than in the dijet or $t\bar t$ channel, but $3\sigma$ significance will require almost $10^3$ fb$^{-1}$ of luminosity.

\section{Squarkonium and other QCD bound states \label{sec-other-QCD-bs}}

So far we discussed the bound states of color-fundamental fermions (toponium) and color-adjoint fermions (gluinonium). We have seen that relative to the decay rate of the tops, the annihilation decays of toponium have a too small branching ratio to be detectable. On the other hand, the gluino is sufficiently long-lived in certain regions of parameter space, so gluinonium may have observable decays to dijets and $t\bar{t}$. Let us now discuss the other possible bound states of particles
charged under QCD. In the MSSM, the only other colored particles are squarks, which are scalars in the fundamental representation. There exist strong production processes (see for example~\cite{Dawson:1983fw,Beenakker:1996ch}) for a squark-antisquark pair
\be
gg \to \sq_i \bar\sq_i\,,\qq
q_i \bar q_j \to \sq_k \bar\sq_l\,,
\ee
a pair of squarks
\be
q_i q_j \to \sq_i \sq_j\,,
\label{di-squark}
\ee
or a squark and a gluino
\be
q_i g \to \sq_i \go \,.
\label{squark-gluino}
\ee
Let us discuss the possible bound states of these particles.

\subsection{Squarkonium}
Squarkonium is a bound state of a squark and an antisquark. As in the case of toponium, the color decomposes as
\be
\mathbf{3} \otimes \mathbf{\bar 3} = \mathbf{1} \oplus \mathbf{8} \,,
\ee
and only the color singlet can bind. The color factor for the singlet bound state is
\be
C_\mathbf{1} = \frac{4}{3} \,.
\ee

In most scenarios squarkonia do not form because the squarks decay too fast. For the gluino, we could make the
single-particle decay rate arbitrarily small by making all the squarks heavy. However, because
squarks have electroweak quantum numbers and are heavier than the lightest supersymmetric particle (LSP),
they can generically decay as
\be
\sq \to \sq'\, W\qq
\mbox{or}\qq
\sq \to q\, \chi \,,
\ee
where $\chi$ is a chargino or neutralino. The rate for these decays is in general comparable to the binding energy
\be
E_b = \frac{4}{9}\bar\alpha_s^2 m_\sq
\ee
of the would-be squarkonium and we obtain a situation similar to the toponium in which the bound state is very broad and it decays primarily by single-particle decays rather than the annihilation decays.

\begin{figure}
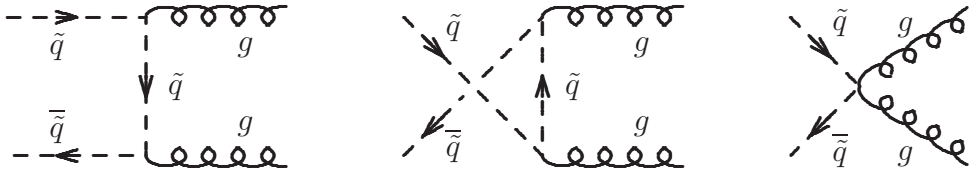

$$\beginpicture
\setcoordinatesystem units <0.75\tdim,0.75\tdim>
\stpltsmbl
\setdashes
\plot -70 35 0 35 0 -35 -70 -35 /
\setsolid
\barrow from -45 35 to -35 35
\barrow from -35 -35 to -45 -35
\barrow from 0 5 to 0 -5
\ellipticalarc axes ratio 2:1 -220 degrees from 0 35 center at 10 35
\ellipticalarc axes ratio 2:1 -280 degrees from 17 31 center at 24 35
\ellipticalarc axes ratio 2:1 -280 degrees from 32 31 center at 39 35
\ellipticalarc axes ratio 2:1 -280 degrees from 47 31 center at 54 35
\ellipticalarc axes ratio 2:1 -150 degrees from 62 31 center at 69 35
\ellipticalarc axes ratio 2:1 220 degrees from 0 -35 center at 10 -35
\ellipticalarc axes ratio 2:1 280 degrees from 17 -31 center at 24 -35
\ellipticalarc axes ratio 2:1 280 degrees from 32 -31 center at 39 -35
\ellipticalarc axes ratio 2:1 280 degrees from 47 -31 center at 54 -35
\ellipticalarc axes ratio 2:1 150 degrees from 62 -31 center at 69 -35
\put {$\sq$} at -45 20
\put {$\bar\sq$} at -45 -20
\put {$\sq$} at 15 0
\put {$g$} at 50 20
\put {$g$} at 50 -20
\linethickness=0pt
\putrule from 0 -60 to 0 60
\putrule from -100 0 to 100 0
\endpicture
\beginpicture
\setcoordinatesystem units <0.75\tdim,0.75\tdim>
\stpltsmbl
\setdashes
\plot -70 35 0 -35 0 35 /
\plot -70 -35 -40 -5 /
\plot -30 5 0 35 /
\setsolid
\barrow from -60 25 to -50 15
\barrow from 0 -5 to 0 5
\barrow from -50 -15 to -60 -25
\ellipticalarc axes ratio 2:1 -220 degrees from 0 35 center at 10 35
\ellipticalarc axes ratio 2:1 -280 degrees from 17 31 center at 24 35
\ellipticalarc axes ratio 2:1 -280 degrees from 32 31 center at 39 35
\ellipticalarc axes ratio 2:1 -280 degrees from 47 31 center at 54 35
\ellipticalarc axes ratio 2:1 -150 degrees from 62 31 center at 69 35
\ellipticalarc axes ratio 2:1 220 degrees from 0 -35 center at 10 -35
\ellipticalarc axes ratio 2:1 280 degrees from 17 -31 center at 24 -35
\ellipticalarc axes ratio 2:1 280 degrees from 32 -31 center at 39 -35
\ellipticalarc axes ratio 2:1 280 degrees from 47 -31 center at 54 -35
\ellipticalarc axes ratio 2:1 150 degrees from 62 -31 center at 69 -35
\put {$\sq$} at -45 30
\put {$\bar\sq$} at -45 -30
\put {$\sq$} at 15 0
\put {$g$} at 50 20
\put {$g$} at 50 -20
\linethickness=0pt
\putrule from 0 -60 to 0 60
\putrule from -100 0 to 100 0
\endpicture
\beginpicture
\setcoordinatesystem units <0.75\tdim,0.75\tdim>
\stpltsmbl
\setdashes
\plot -35 35 0 0 -35 -35 /
\setsolid
\barrow from -25 25 to -15 15
\barrow from -15 -15 to -25 -25
\startrotation by 0.8 0.5 about 0 0
\ellipticalarc axes ratio 2:1 -220 degrees from 0   0 center at 10 0
\ellipticalarc axes ratio 2:1 -280 degrees from 17 -4 center at 24 0
\ellipticalarc axes ratio 2:1 -280 degrees from 32 -4 center at 39 0
\ellipticalarc axes ratio 2:1 -280 degrees from 47 -4 center at 54 0
\ellipticalarc axes ratio 2:1 -150 degrees from 62 -4 center at 69 0
\stoprotation
\startrotation by 0.8 -0.5 about 0 0
\ellipticalarc axes ratio 2:1 220 degrees from 0  0 center at 10 0
\ellipticalarc axes ratio 2:1 280 degrees from 17 4 center at 24 0
\ellipticalarc axes ratio 2:1 280 degrees from 32 4 center at 39 0
\ellipticalarc axes ratio 2:1 280 degrees from 47 4 center at 54 0
\ellipticalarc axes ratio 2:1 150 degrees from 62 4 center at 69 0
\stoprotation
\put {$\sq$} at -10 32
\put {$\bar\sq$} at -10 -32
\put {$g$} at 23 30
\put {$g$} at 23 -35
\linethickness=0pt
\putrule from 0 -60 to 0 60
\putrule from -60 0 to 60 0
\endpicture$$
\caption{Diagrams for the dominant annihilation decay of stoponium. The first two diagrams do not actually contribute because they vanish at threshold.}
\label{fig-sqsq-annih}
\end{figure}

\begin{table}
$$\begin{array}{|l|c|c|c|c|}
\hline
\,\mbox{Model} &\, m_\st \mbox{ (GeV)} \,&\, E_b \mbox{ (GeV)} \,&\, 2\,\Gamma_\st \mbox{ (GeV)} \,&\, \Gamma^{\rm ann} \mbox{ (GeV)} \\
\hline
\,\mbox{SPS 5 (mSUGRA with light stop)} \,&\, 262   \,&\,    2.3 \,&\, 0.087 \,&\, 0.0028 \\
\,\mbox{All other SPS points}               \,&\, \geq 400 \,&\, \sim 5 \,&\, \geq 3 \,&\, \sim 0.004 \\
\hline
\end{array}$$
\caption{The mass of the lighter stop $m_\st$, the binding energy $E_b$, twice the stop decay rate $2\Gamma_\st$, and the stoponium annihilation rate $\Gamma^{\rm ann}$, for the SPS benchmark points~\cite{Allanach:2002nj}. In all the SPS scenarios, the stop decays roughly equally via $\st \to \bar\chi^0 t$ and $\st \to \bar\chi^+b$ (only the latter is present in SPS 5).}
\label{tab-stst-SPS}
\end{table}

The only possible exception is stoponium. There, the stop can be lighter than the top plus LSP, so the weak decays are forbidden.
This is not the case in most of the parameter space of the mSUGRA ansatz. In SPS 5, even though the width is smaller than the binding energy, single-stop decays still dominate over the annihilation decays (see Table~\ref{tab-stst-SPS}). However, the situation is much more favorable in certain other motivated scenarios discussed in more detail in~\cite{Martin:2008sv,Herrero:1987df}.

Stoponium ($J^{PC}=0^{++}$) is produced primarily by gluon fusion. Its annihilation into $gg$ via the diagrams in Figure~\ref{fig-sqsq-annih} has the rate
\be
\Gamma(\mathbf{1}\to gg) = \frac{4\pi\alpha_s^2}{3m_\sq^2}\l|\psi(\vv{0})\r|^2 = \frac{32}{81}\alpha_s^2\bar\alpha_s^3 m_\sq \,.
\label{stst-annih}
\ee
Unfortunately, this is about $200$ times smaller than for gluinonia of the same mass mostly due to the different color factors (compare to (\ref{1annih}), and note that in the gluinonium case there are also contributions from the octets), and twice smaller than for a heavy quarkonium of the same mass (see (\ref{ttbar2gg})), and the production cross section, which is given by (\ref{hadron-cs-1}), is correspondingly smaller. As a result, the stoponium cannot be seen in the dijet or $t\bar{t}$ channels even when the annihilation decays dominate.

However, since the squarks are charged, the annihilation of stoponium to $\gamma\gamma$ is relevant.
Although the branching ratio to  $\gamma\gamma$
is only $\sim(8/9)(\alpha/\alpha_s)^2 \simeq 0.005$, for certain regions of MSSM
parameter space the cross section may still be of order 1 fb which makes it possibly observable. Other stoponium decay
modes including $ZZ$ and $W^+W^-$ could possibly be visible as well. Stoponium decays
have been studied in detail in~\cite{Drees:1993yr,Drees:1993uw,Martin:2008sv,Martin:2009dj,Younkin:2009zn} (note that~\cite{Drees:1993yr,Drees:1993uw} contain certain factor-of-$2$ errors -- see~\cite{Moxhay:1985bg,Gorbunov:2000tr,Martin:2008sv}).

\subsection{Di-squark and squark-gluino bound states}
Instead of a squark and anti-squark, another possible bound state might be two squarks or a squark and a gluino.
For di-squarks, the relevant color decomposition is
\be
\mathbf{3} \otimes \mathbf{3} = \mathbf{\bar 3} \oplus \mathbf{6}
\ee
and the color factors are
\be
C_\mathbf{\bar 3} = \frac{2}{3}\,,\qq
C_\mathbf{6} = -\frac{1}{3} \,,
\ee
so the triplet has an attractive potential and can potentially form a bound state.
For a squark-gluino pair, the decomposition is
\be
\mathbf{3} \otimes \mathbf{8} = \mathbf{3} \oplus \mathbf{\bar 6} \oplus \mathbf{15}
\ee
and
\be
C_\mathbf{3} = \frac{3}{2}\,,\qq
C_\mathbf{\bar 6} = \frac{1}{2}\,,\qq
C_\mathbf{15} = -\frac{1}{2} \,,
\ee
which means that the potential for the triplet and the sextet is attractive and could potentially support spin-$\frac{1}{2}$ bound states of these two particles.

However, the di-squarks and the squark-gluino bound states suffer from the same problem as the squarkonium, namely the too rapid decays of the squarks. Furthermore, even the special cases in which the lighter stop is sufficiently stable are not available here because, unlike in the case of the stoponium, producing a di-stop or a stop-gluino bound state via (\ref{di-squark}) or (\ref{squark-gluino}), respectively, would require having top quarks as incoming partons.

\subsection{Other QCD bound states}
There are many more possible bound states of hypothetical new colored particles.
For example, superheavy quarkonia have been studied in~\cite{Barger:1987xg,Arik:2002nd} and bound states of $SU(2)_L$-doublet color-octet scalars in~\cite{Kim:2008bx}. It would also be interesting look at bound states of Kaluza-Klein (KK) excitations of quarks and gluons in theories with extra dimensions.
For example, Universal Extra Dimension (UED) models have TeV-scale flat extra dimensions
 with all the standard model fields propagating in the bulk, so they all have massive excitations.
For certain regions of parameter space KK quarks can be sufficiently stable to form bound states~\cite{Carone:2003ms,Fabiano:2008xk}.
Even though the KK particles are color adjoints or fundamentals, like the gluino or squarks we have already considered,
the details of the phenomenology will be somewhat different because of the different spins as well as the different range of possibilities for the single-particle lifetimes, and it would be interesting to study this in more detail in a future work. If annihilation decays can be detected at the LHC, one interesting application would be to try to use them to distinguish extra dimensions from
supersymmetry. From Figure~\ref{fig-pT-angle}, there is clearly some spin information contained in the dijets from annihilation,
but isolating a clean sample to see this effect may be impossible in practice. The size of the cross sections for the different annihilation channels might also be a useful discriminator, when combined with other information.

\section{Bound states with new forces\label{sec-bs-new-forces}}
The previous sections have all looked at bound states for which the binding force is the strong force of QCD.
We have seen that being charged under the strong force allows particles to be produced with substantial cross sections, and
substantial binding and annihilation rates, as high as tens of picobarns.
Our discussion of the phenomenology of colored bound states was centered around supersymmetry, in particular, the MSSM,
 which has new particles but no new forces as compared to the standard model.
However, there are plenty of models which do have new forces. Technicolor is an obvious example, but even within supersymmetric
models, dynamical supersymmetry breaking often involves something beyond $SU(3)\times SU(2)\times U(1)$. It is certainly possible
for there to be a new force, even stronger than QCD, which would lead to quite substantial annihilation decay rates.

Let us suppose there is a new force associated with a new $SU(N)$ gauge group.
 If the associated strong coupling scale $\Lambda_N$ is sufficiently
large (say 10 TeV), then all the particles produced at the LHC would be confined into $SU(N)$-singlets.
 This case is not interesting for the present paper, as there
is no way to distinguish particles bound by this force from fundamental resonances. Instead, we are interested in a situation in which the LHC will be able to produce particles with mass $m \gg \Lambda_N \gtrsim \Lambda_{\mathrm{QCD}}$. We further assume they are charged under both QCD and $SU(N)$, so that they will also have substantial production rates. Recall that the annihilation rates are proportional to $\l|\psi(\vv{0})\r|^2 \propto \bar \alpha_s^3$. Since these rates scale as the third power of the coupling, the new gauge coupling does not have to be much larger than the QCD coupling to drastically increase the rate.

To distinguish annihilation decays from simple resonances, we also need to
 have some other way to observe these new particles. For example, this may be through single particle decays, as with the gluino.
In fact, there are many models in which weird things happen when there are new forces, such as displaced vertices in hidden valley models~\cite{Strassler:2006im}, or macroscopic strings in quirks~\cite{Kang:2008ea}. We will focus only on annihilation decays, leaving their coordination with the rich field of new-force model signatures for future consideration.

\subsection{Increasing the annihilation rate\label{sec-annih-rate}}
For gluinos of phenomenologically relevant masses,
we saw that as the gluino becomes more stable the gluinonium annihilation cross section saturates at a few percent of the continuum gluino pair production cross section. Let us discuss which factors set the scale for this number and what can make it larger, and then consider the situation in which the binding particles have an additional attractive force, beyond the $SU(3)$ of QCD.

The bound state production cross section is given by
\be
\sigma^{\rm bound} \simeq
\frac{\mathcal{L}(4m^2)}{m} \int \hat\sigma^{\rm bound}(\hat s)\, \rd\sqrt{\hat s}
\sim \frac{\mathcal{L}(4m^2)}{m}\,\frac{\alpha_s^2}{m^4}\l|\psi(\vv{0})\r|^2
\sim \mathcal{L}(4m^2)\frac{\alpha_s^2 C^3\bar\alpha_s^3}{m^2}\,,
\ee
where $\mathcal{L}(M^2)$ is the parton luminosity,  Eq.~\eqref{lumi}.
If we neglect the dependence of the parton luminosities on the energy,
then the parton-level continuum production cross section $\hat\sigma^{\rm cont}(\hat s) \sim \alpha_s^2/\hat s$ gives
\be
\sigma^{\rm cont} \sim \frac{\mathcal{L}(4m^2)}{m}\int \hat\sigma^{\rm cont}(\hat s)\, \rd\sqrt{\hat s}
\sim \mathcal{L}(4m^2)\,\frac{\alpha_s^2}{m^2} \,,
\ee
so that
\be
\frac{\sigma^{\rm bound}}{\sigma^{\rm cont}} \sim C^3\bar\alpha_s^3 \,.
\label{bound-cont-ratio}
\ee
There are additional factors coming from the fact that only part of the color configurations bind,
and the fall off of the parton luminosities with energy, but this qualitative scaling still holds.
The point is that because $C$ and $\bar\alpha_s$ are raised to the third power in this relation, the fraction of annihilation decays
can be significantly enhanced over the continuum if the coupling or color factors are larger than they are for QCD.

It is interesting to also consider how large the fraction of annihilation decays can be, while the binding force is still perturbative.
When the coupling $\bar\alpha_s$ is increased, it not only increases the binding rate, but also adds a radiative correction to the
continuum production. This can lead to a significant Sommerfeld enhancement of the total cross section.
From Eq.~(\ref{resummation}), the additive correction to the continuum production near the threshold is roughly
\be
\Delta\hat\sigma^{\rm cont}(\hat s) \sim \frac{\alpha_s^2}{m^2}\,C\bar\alpha_s \,.
\ee
This extra contribution needs to be integrated up to roughly the binding energy $E_b \sim C^2\bar\alpha_s^2 m$. This gives
\be
\frac{\Delta\sigma^{\rm cont}}{\sigma^{\rm cont}}
\sim \frac{1}{\sigma^{\rm cont}}\int_{2m}^{2m + E_b} \Delta\hat\sigma^{\rm cont}(\hat s)\, \rd\sqrt{\hat s}
\sim C^3\bar\alpha_s^3
\sim \frac{\sigma^{\rm bound}}{\sigma^{\rm cont}} \,.
\ee
Thus for large $C\bar\alpha_s$ the continuum rate will also be proportionally larger, and there will never be more than $\sim 50\%$ annihilation decays (with  $\bar\alpha_s\lesssim 1$).

In the more extreme limit in which the theory becomes strongly coupled and confining, the bound states will look just like simple resonances decaying either back into the standard model or within the strongly-coupled sector itself. But, in the regime of interest, when $\bar\alpha_s$ is perturbative but $(C\bar\alpha_s)^3$ is larger than for gluinonium, the binding particles will be observable on their own and the annihilation decays of their bound states will be significantly enhanced.
In the next subsection we exemplify this with a toy model in which the particles are charged under an additional force.

\subsection{Example model\label{sec-new-forces-example}}

Suppose there exist color adjoint fermions, like the gluino, which also transform under some representation $R$ of a new $SU(N)$ with a coupling constant $\alpha_N$.
Their pair production processes will be like for the gluino, but without the contribution from virtual squarks. For $q\bar{q}$ initial states, only the anti-symmetric color adjoint state $\mathbf{8_A}$ is produced near the threshold. For $gg$ initial states, pairs in the $\mathbf{1}$, $\mathbf{8_S}$ and $\mathbf{27}$ color configurations will be produced with relative cross sections $\hat\sigma_0(\mathbf{1}):\hat\sigma_0(\mathbf{8_S}):\hat\sigma_0(\mathbf{27}) = 1:2:3$~\cite{Hagiwara:2009hq}. Since the initial states are $SU(N)$ singlets, only $SU(N)$-singlet configurations can be produced. Because of the sum over final states the rate is enhanced by an additional factor of the dimension of the representation, $D_R$.

For bound states, the cross-sections will also include the appropriate $|\psi(\vv{0})|^2$ factor for each color configuration. For computing those, note that the couplings in the potentials (\ref{V(r)}) are given by the simple replacements
\be
C\,\bar\alpha_s \to C\,\bar\alpha_s + C_R\bar\alpha_N \,,
\ee
where $C$ is the relevant QCD color factor from (\ref{gluinonium-color-factors}) for the $\mathbf{1}$, $\mathbf{8_S}$, $\mathbf{8_A}$ or $\mathbf{27}$ states, and $C_R$ is the quadratic Casimir of the representation $R$ of $SU(N)$. For example, for the fundamental representation of $SU(N)$, $D_R = N$ and $C_R = (N^2-1)/2N$, and for the adjoint representation $D_R = N^2-1$ and $C_R = N$.

To be explicit, the cross sections for the pseudoscalar bound states in the $(\mathbf{1}, \mathbf{1})$, $(\mathbf{8_S}, \mathbf{1})$ and $(\mathbf{27}, \mathbf{1})$ representations of $SU(3)_{\rm QCD} \times SU(N)$ will be like the cross section of the $\mathbf{1}$ gluinonium of the MSSM times the factors
\be
D_R \l(\frac{3\,\bar\alpha_s + C_R\,\bar\alpha_N}{3\,\bar\alpha_s}\r)^3\,, \q
2D_R \l(\frac{3/2\,\bar\alpha_s + C_R\,\bar\alpha_N}{3\,\bar\alpha_s}\r)^3\,, \q
3D_R \l(\frac{-\bar\alpha_s + C_R\,\bar\alpha_N}{3\,\bar\alpha_s}\r)^3 \,,
\ee
respectively. The $(\mathbf{8_A}, \mathbf{1})$ vector will be like the $\mathbf{8_A}$ gluinonium times
\be
D_R\l(\frac{3/2\,\bar\alpha_s + C_R\,\bar\alpha_N}{3/2\,\bar\alpha_s}\r)^3 \,.
\ee
Notice that even in relatively small representations and with order $1$ values of $\bar\alpha_N/\bar\alpha_s$ the annihilation decay rates will be orders of magnitude larger than anything in the MSSM.

\begin{figure}[t]
\begin{center}
\psfrag{F}[][][0.8]{$5\sigma$}
\psfrag{G}[][][0.8]{$95\%$}
\psfrag{H1}[][][0.8]{$SU(4)$ adjoint}
\psfrag{H2}[][][0.8]{$\bar\alpha_N = 3\bar\alpha_s$}
\psfrag{I1}[][][0.8]{$SU(4)$ adjoint}
\psfrag{I2}[][][0.8]{$\bar\alpha_N = \bar\alpha_s$}
\psfrag{J1}[][][0.8]{$SU(3)$ fundamental}
\psfrag{J2}[][][0.8]{$\bar\alpha_N = \bar\alpha_s$}
\psfrag{K}[][][0.8]{gluinonium}
\includegraphics[width=0.8\textwidth]{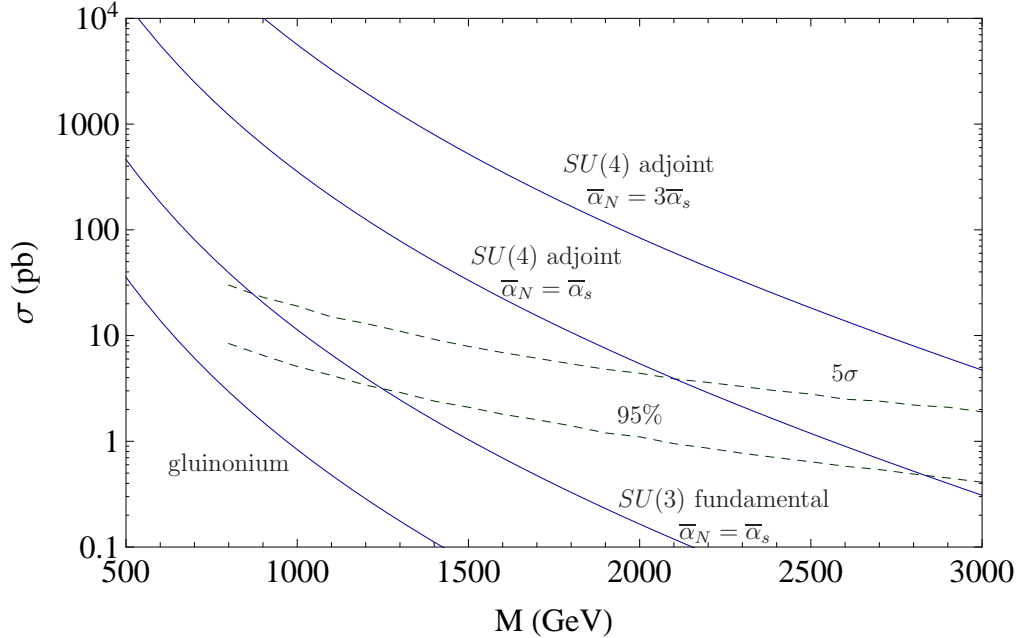}
\end{center}
\caption{Dijet search reach ($95\%$ CL exclusion and $5\sigma$ discovery) with 100 pb${}^{-1}$ at {\sc CMS}~\cite{CMS-NOTE-2006/070}, the MSSM gluinonium signal, and several models in which the particles are bound also by new forces.}
\label{fig-newforces}
\end{figure}

Figure~\ref{fig-newforces} compares the dijet rates through annihilation decay to the {\sc CMS} dijet search reach for several toy models, including the one with the MSSM gluino. Even when the additional force has the same strength as QCD ($\bar\alpha_N = \bar\alpha_s$), the cross section is enhanced by more than a factor of $10$ if our ``gluinos'' are charged under the fundamental representation of a new $SU(3)$ group. If they are charged under the adjoint of a new $SU(4)$, the enhancement is almost $1000$. If the new force is somewhat stronger, $\bar\alpha_N = 3\bar\alpha_s$ (which corresponds to a confinement scale of about $\Lambda_N \sim 10$ GeV, assuming for simplicity the same running as in QCD), the signal will be $10000$ times larger than the annihilation decay signal from MSSM gluinonia.

The bound states may also annihilate into hidden sector dijets. But note that
if the QCD and hidden sector couplings are comparable, the branching ratio into
QCD dijets will still be of order 1, and the dijet rate will still have an enormous
enhancement over annihilation decay rates in the MSSM. The phenomenology of
decays into the hidden sector is very interesting, but model dependent,
so we will not consider it here. In summary, there is plenty of parameter space for complete models of this kind which are consistent with the various experimental bounds, in which such annihilation decays can provide spectacular signals at the LHC.

\section{Summary and conclusions\label{sec-conclusions}}
In this paper we have considered the annihilation decays of a variety of possible bound state particles at the LHC.
A number of examples were discussed at length, including bound states of top quarks, toponium, and bound states of
gluinos, gluinonium. We also discussed bound states of squarks, squarkonium, and bound states associated with possible new strong
forces.
We have discussed how the spins and color representations of the particles determine the possible bound states,
what determines their production channels and their cross sections, and how the mass spectrum
of the model determines whether the bound states form and whether they
decay by annihilation or through single-particle decays. Whenever the observation of the annihilation signal
(or its absence) is feasible, it can provide valuable direct or indirect information about the
underlying particles.

 In the first and only standard model case, toponium, we found that the width of the top quark itself precludes
any reasonable hope of finding toponium. Not only does the top's relatively large width force the top quark to decay much faster than it can annihilate, but it even decays too fast for a well-defined bound state to form, broadening the annihilation decay spectrum. There is no sharp distinction between the dijets coming from toponium
annihilation decays and the contribution that virtual top quarks give to continuum dijet production. One might
have also hoped to find evidence for toponium from the detailed spectrum of the weak decay products, but since
there is no sharp feature at the $t\bar{t}$ spectrum, the resolution required is beyond the ability of the LHC.

Although the top lifetime is fixed, gluinos may live long enough for the gluinonium to decay predominantly by annihilation in certain regions of
parameter space, including some of the standard SPS points. We found that gluinonia may have substantial decay rates both
to dijets and to $t\bar{t}$, and that both of these decay modes, although difficult to see, may be detectable at the LHC if the gluino mass is of the order
of several hundred GeV.
Looking at the mass distributions we found that
both channels could provide peaks for mass determination. Moreover, the angular distributions in
 the dijet channel may help determine the spins of the resonances and thereby put constraints on the properties of the binding
particle. However, due to the low signal-to-background ratio it will be challenging to extract spin information from the data.

For the case of bound squarks, squarkonium, observing annihilation decays into either dijets or $t\bar t$ pairs is
practically impossible.
First of all,  despite the fact that the production is through gluon fusion, squarkonia cross sections are always
several orders of magnitude smaller than gluinonia cross sections because of color factors.
Moreover, for generic models, the squarks, like the top, will decay too fast relative to the rate of the annihilation processes or even before they form bound states, and the large width will smear the bound state resonance signal.
Only stops may have a long enough lifetime so that the  annihilation decay signal might possibly be seen.
For these the best hope is using the small branching ratio to annihilate into two photons. Although the cross section
for these decays is less than 1 fb for stop masses above 200 GeV, this channel may be feasible for high luminosity. In fact, this channel might even be feasible for the gluinonium because of its large color factors even though the gluinos can annihilate into two photons only through loop diagrams.
We also considered bound states of a squark and a gluino and two squarks (as opposed to a squark-anti-squark). Neither of these
have visible annihilation decay signals.

Finally, we looked at annihilation decay cross sections in models with new forces. With a new confining force, the rate
for annihilation decays may easily be much larger than in the supersymmetric models which only bind with QCD. There is a huge
space of possible models, and just restricting to one simple class, we found that the signals for decays of these bound states
are easy to see. Of course, in the limit that the new force is infinitely strong, the annihilation decays are just decays of
a new resonance. However, in intermediate regimes it is easy to have both annihilation decays and other signatures, so that
the annihilation decays provide complementary information.

Overall, it seems that annihilation decays of particles bound through QCD are difficult, but possible, to see,
while particles bound through stronger forces will show up quickly. If the binding
is not too strong, as with QCD, we will almost certainly have information about other decays of the constituent particles
before we find the bound states. This will be useful in directing our search -- knowing the mass ahead of time can improve
the search strategy, especially in the dijet channel.
Conversely, if the binding is very strong, exploring what the resonance may be a bound state of
may help direct searches for other physics associated with the binding particles. In general, annihilation decays
of bound states have the potential to provide an additional handle on the new physics we are likely to see at the LHC.

\section*{Acknowledgments}
We thank Andrea De Simone for collaboration on part of this work and Hiroshi Yokoya for helpful discussions.
Some of the computations in this paper were performed on the Odyssey cluster supported by the FAS Research Computing Group
at Harvard University. Our research is supported in part by the National Science Foundation under grant PHY-0804450
and by the Department of Energy OJI program, under Grant DE-AC02-76CH03000.

\appendix

\section{Bound state formalism\label{sec-gen-bound}}

For computing the bound-state effects, as well as finite-width effects, it is useful to consider the Green's function $G(\vv{x},E)$ of the Schr\"{o}dinger equation describing the two-particle system:\footnote{Here $\vv{x}$ is the distance between the particles and we will be interested in $G(\vv{x=0},E)$ because we want to create the particles at zero separation and annihilate them at zero separation at a later time (the time is represented by its Fourier variable $E$).}
\be
\l[-\frac{\nabla^2}{m} + V(r) - E\r] G(\vv{x},E) = \delta^{(3)}(\vv{x}) \,,
\ee
where $E = \sqrt{\hat s} - 2m$. For zero angular momentum, the solution is~\cite{Fadin:1988fn}
\bea
G(\vv{0},E)
&=& -\frac{m^2}{4\pi}\l[\sqrt{-\frac{E}{m}} - C\bar\alpha_s\l[\ln\l(\frac{|C|\bar\alpha_s}{2}\sqrt{-\frac{m}{E}}\r) - \digamma\l(1-\frac{C\bar\alpha_s}{2}\sqrt{-\frac{m}{E}}\r) - \gamma_E \r]\r] \nn\\
&=& -\frac{m^2}{4\pi}\l[\sqrt{-\frac{E}{m}} - C\bar\alpha_s\ln\l(\frac{|C|\bar\alpha_s}{2}\sqrt{-\frac{m}{E}}\r) - \frac{2}{\sqrt m}\sum_{n=1}^\infty\frac{E_n}{\sqrt{-E} - \mbox{sign}(C)\sqrt{E_n}}\r]
\label{Green-soln}
\eea
where $\digamma(x) = \Gamma'(x)/\Gamma(x)$ is the digamma function, $\gamma_E = -\digamma(1) \simeq 0.577$ is Euler's constant, $C$ is the color factor (\ref{color-factor}) and $E_n = C^2 \bar\alpha_s^2 m/4n^2$ are the energies of the radial excitations. The solution describes both the bound states and the near-threshold continuum. Particles with width $\Gamma$ (which gives width $2\Gamma$ to the bound states) are described by $E \to E + i\Gamma$~\cite{Fadin:1988fn,Strassler:1990nw}.

\subsection{Production cross section}

By the optical theorem, the production cross section will be proportional to $\mbox{Im}\,G(\vv{0},E)$, thus it can be written as~\cite{Fadin:1987wz,Fadin:1987mj,Fadin:1988fn,Fadin:1990wx,Strassler:1990nw} \be
\hat\sigma(E) = \hat\sigma_0(E)\,
\frac{\mbox{Im}\,G(\vv{0},E)}{\mbox{Im}\,G_0(\vv{0},E)} \,,
\label{ImG-correction}
\ee
where $\hat\sigma_0$ is production cross section of the pair of particles (in the particular angular momentum and color representation) without the binding corrections and $G_0$ is the free Green's function, that is (\ref{Green-soln}) with $C = 0$ (and $\Gamma = 0$), namely
\be
G_0(\vv{0},E) = -\frac{m^2}{4\pi}\sqrt{-\frac{E}{m}} \,, \qqq
\mbox{Im}\,G_0(\vv{0},E) = \frac{m^2}{4\pi}\sqrt{\frac{E}{m}}\,\theta(E)
\simeq \frac{m^2}{4\pi}\beta\,\theta(E) \,,
\label{Green-free}
\ee
where $\beta \equiv \sqrt{1-4m^2/\hat s}$ is the velocity of the particles in their center-of-mass frame. Then
\be
\hat\sigma(E) = \frac{4\pi}{m^2}\frac{\hat\sigma_0(E)}{\beta}\,\mbox{Im}\,G(\vv{0},E)
= \frac{\hat\sigma_0(E)}{2m^2\,\Phi_2(E)}\, \mbox{Im}\,G(\vv{0},E) \,,
\label{sigma-ImG/Phi2}
\ee
where we wrote the last expression in terms of the 2-particle phase space
\be
\Phi_2 = \int\frac{\rd^3p_1}{(2\pi)^3\,2p_1^0} \int\frac{\rd^3p_2}{(2\pi)^3\,2p_2^0}\,
(2\pi)^4\delta^3\l(\mathbf{p}_2 - \mathbf{p}_1\r)
\delta\l(m\beta^2 - \frac{\mathbf{p}_1^2}{2m} - \frac{\mathbf{p}_2^2}{2m}\r)
= \frac{\beta}{8\pi}
\label{Phi2}
\ee
since the combination $\hat\sigma_0/\Phi_2$ remains well-behaved as we go below threshold.

For an attractive potential ($C > 0$), the last term of (\ref{Green-soln}) grows large near the bound state energies $E = -E_n$ if $E_n \gg \Gamma$:
\be
G(\vv{0},E) \simeq -\sum_n \frac{|\psi_n(\vv{0})|^2}{E + E_n + i\Gamma}
\simeq -\sum_n\frac{2M_n|\psi_n(\vv{0})|^2}{\hat s - M_n^2 + 2iM_n\Gamma} \,,
\label{G0-near-bs}
\ee
\be
\mbox{Im}\,G(\vv{0},E)
\simeq \sum_n |\psi_n(\vv{0})|^2\frac{\Gamma}{(E+E_n)^2 + \Gamma^2}
\simeq \sum_n |\psi_n(\vv{0})|^2\frac{4M_n^2\Gamma}{(\hat s - M_n^2)^2 + 4M_n^2\Gamma^2} \,,
\label{ImGreen-bound-narrow}
\ee
where $|\psi_n(\vv{0})|^2 = C^3\bar\alpha_s^3m^3/(8\pi n^3)$ refers to the hydrogen-like wavefunctions describing the bound states and $M_n = 2m - E_n$ are their masses. In the limit $\Gamma\to0$,
\be
\mbox{Im}\,G(\vv{0},E) = \pi\sum_n |\psi_n(\vv{0})|^2\,\delta(E+E_n)
= \sum_n M_n |\psi_n(\vv{0})|^2\,\Phi_1 \,,
\label{ImGreen-bound-zero}
\ee
\be
\hat\sigma(E) = \frac{8\pi}{m}\frac{\hat\sigma_0(E)}{\beta} \sum_n|\psi_n(\vv{0})|^2\,\Phi_1
= \sum_n\frac{|\psi_n(\vv{0})|^2}{m}\,\hat\sigma_0(E)\, \frac{\Phi_1}{\Phi_2} \,,
\label{sigma-Phi1/Phi2}
\ee
where $\Phi_1 = 2\pi\,\delta(\hat s-M_n^2)$ is the usual single-particle phase space. The last expression makes the agreement of the whole formalism with (\ref{times-wavefunction}) manifest.\footnote{In the case that the two particles are identical, $\Phi_2$ in (\ref{Phi2}) will actually be twice the phase space. This is consistent with the factor of $1/2$ discussed after (\ref{identical-factor}).} However, (\ref{times-wavefunction}) is more general since it holds also for non-Coulombic potentials (but it is valid only for narrow bound states).

More generally, the imaginary part of (\ref{Green-soln}) is~\cite{Fadin:1987wz,Fadin:1988fn,Fadin:1990wx}
\be
\mbox{Im}\,G(\vv{0},E) = \frac{m^2}{4\pi}\l[v_+ + C\bar\alpha_s \tan^{-1}\l(\frac{v_+}{v_-}\r)
+ \sum_{n=1}^\infty E_n
\frac{C\bar\alpha_s \Gamma/n + 2v_+\l(\sqrt{E^2 + \Gamma^2} + E_n\r)}{\l(E+E_n\r)^2 + \Gamma^2}\r] \,, \label{ImGreen}
\ee
where
\be
v_\pm = \sqrt{\frac{\sqrt{E^2+\Gamma^2}\pm E}{2m}} \,.
\ee
To see what (\ref{ImGreen}) means, suppose $\Gamma\to 0$. For $E < 0$, Eq. (\ref{ImGreen}) reduces to a sum of narrow Breit-Wigners (\ref{ImGreen-bound-narrow}) in the attractive case and vanishes in the repulsive case. For $E > 0$:
\be
\mbox{Im}\,G(\vv{0},E) = \frac{C\bar\alpha_s m^2}{4}\, \frac{1}{1 - \exp\l(-\pi C\bar\alpha_s/\beta\r)} \,.
\label{resummation}
\ee
For a repulsive potential, the imaginary part (\ref{resummation}) is suppressed to zero at the na\"{\i}ve threshold $E = 0$, but for an attractive potential it starts with a finite value
\be
\mbox{Im}\,G(\vv{0},E\to 0^+) = \frac{C\bar\alpha_s m^2}{4} \,.
\label{ImG-thresh}
\ee
at the threshold.

\subsection{Annihilation processes}

Consider the process
\be
ab \to \alpha\beta \to AB \,.
\label{abABcd}
\ee
When $\alpha\beta$ form a narrow bound state, the standard resonance production formula (\ref{sigma-hat-general-partial}) gives
\be
\hat\sigma_{ab\to (\alpha\beta)\to AB}(\hat s) = \frac{4\pi \l(2J+1\r) D_{(\alpha\beta)}}{D_a D_b}\frac{\Gamma_{(\alpha\beta)\to ab}\,\Gamma_{(\alpha\beta)\to AB}}{\l(\hat s-M^2\r)^2 + \hat s\,\Gamma^2}\qq \l[\;\times 2 \q{\rm if}\q a = b\;\r] \,,
\label{resonance-cs}
\ee
where $M \simeq 2m$ is the mass and $J$ is the spin of the bound state and $D_x$ denotes the dimension of the color representation of the particle $x$.

When the bound state is not narrow or we want to obtain the contribution from the near-threshold $\alpha\beta$ continuum as well, we can use the fact that $G(\vv{0},E)$ is the two-point function describing the pair, so the matrix element (squared) is proportional to $|G(\vv{0},E)|^2$, while all the short-distance factors can be taken from the narrow bound state case. Dividing (\ref{resonance-cs}) by the narrow-width $|G(\vv{0},E)|^2$ from (\ref{G0-near-bs})\footnote{More specifically, $|G(\vv{0},E)|^2$ from (\ref{G0-near-bs}) includes the interference with the radial excitations, which we should have, in principle, included in (\ref{resonance-cs}) as well. We ignore this for simplicity of presentation.} and multiplying by the general $|G(\vv{0},E)|^2$ we obtain
\be
\hat\sigma_{ab\to \alpha\beta \to AB}(E)
= \frac{\pi \l(2J+1\r) D_{(\alpha\beta)}}{D_a D_b}\,
\frac{\Gamma_{(\alpha\beta)\to ab}\,\Gamma_{(\alpha\beta)\to AB}}{4m^2|\psi(\vv{0})|^4}\, |G(\vv{0},E)|^2
\q \l[\;\times 2 \;\,{\rm if}\;\, a = b\;\r] \,.
\label{sigma-annih-gen-Gammas}
\ee
Note that $|\psi(\vv{0})|^4$ in the denominator cancels the $|\psi(\vv{0})|^2$ factors in the annihilation rates, so the prefactor of $|G(\vv{0},E)|^2$ in (\ref{sigma-annih-gen-Gammas}) depends only on the short-distance physics. We can see this more explicitly by comparing (\ref{sigma-Phi1/Phi2}) and (\ref{sigma-hat-general-total})\footnote{This expression applies if $A$ and $B$ are massless particles with 2 polarizations, but can be easily generalized.} which allows us to express the bound state annihilation rates in terms of production cross sections of free particles, giving
\be
\hat\sigma_{ab\to \alpha\beta \to AB}(E)
= \frac{D_A D_B}{4\pi \l(2J+1\r) D_{(\alpha\beta)} m^2}\,
\frac{\hat\sigma_0^{ab\to\alpha\beta}(E)}{\Phi_2}\,
\frac{\hat\sigma_0^{AB\to\alpha\beta}(E)}{\Phi_2}\, \l|G(\vv{0},E)\r|^2
\q \l[\;\times \frac{1}{2} \;\,{\rm if}\; A = B\;\r]
\label{sigma-annih-gen-sigmas}
\ee
Written in this form, the expression is useful also for color representations that do not bind.

For the bound states in the narrow-width limit, (\ref{G0-near-bs}) and (\ref{ImGreen-bound-narrow}) show that
\be
\l|G(\vv{0},E)\r|^2 = \frac{|\psi(\vv{0})|^2}{\Gamma}\, \mbox{Im}\,G(\vv{0},E)
\to \frac{|\psi(\vv{0})|^2}{\Gamma_{\rm tot}/2}\, \mbox{Im}\,G(\vv{0},E) \,,
\ee
where in the last step we promoted the decay rate to include not just the constituent particle widths $\Gamma$ but also the (sum of all the) annihilation rates $\Gamma_{\rm ann}$:
\be
\Gamma_{\rm tot} = 2\,\Gamma + \Gamma_{\rm ann} \,.
\label{sum-rates}
\ee
This makes sense because then (\ref{sigma-annih-gen-Gammas}) or (\ref{sigma-annih-gen-sigmas}) reduces to a branching ratio times the production cross section (\ref{sigma-ImG/Phi2}):
\be
\hat\sigma_{ab\to \alpha\beta \to AB}(E)
= \frac{\Gamma_{(\alpha\beta)\to AB}}{\Gamma_{\rm tot}} \cdot
\hat\sigma_0^{ab\to\alpha\beta}(E)\,
\frac{\mbox{Im}\,G(\vv{0},E)}{2m^2\,\Phi_2} \,.
\ee

When the constituent particle width $\Gamma$ is not much smaller than the binding energy $E_b$, $|G(\vv{0},E)|^2$ is no longer proportional to $\mbox{Im}\,G(\vv{0},E)$ so the branching ratio is not simply determined by $\Gamma_{(\alpha\beta)\to AB}/\Gamma_{\rm tot}$. In fact, the annihilation rates are not even well-defined in this situation because the bound states are not well-defined. Nevertheless, we can still use (\ref{sigma-annih-gen-sigmas}).

However, this method cannot be used when the annihilation rate $\Gamma_{\rm ann}$ is the one that becomes comparable to $E_b$, because the width $\Gamma$ in this formalism is the single-particle decay width, without taking into account the possibility of annihilation (see, e.g., the derivations in~\cite{Fadin:1988fn,Strassler:1990nw}). But this regime will not usually be relevant to actual systems because of (\ref{scales}).

Let us now study the behavior in the continuum. Note that for $|E|,\Gamma \lesssim E_b$ we can approximate (\ref{Green-soln}) by
\be
G(\vv{0},E)
\simeq -\frac{C\bar\alpha_s m^2}{4\pi}\l[\pi\cot\l(\frac{\pi C\bar\alpha_s}{2}\sqrt{-\frac{m}{E}}\r) + \gamma_E\r] \,.
\ee
For $E > 0$ this is
\be
G(\vv{0},E)
\simeq -\frac{C\bar\alpha_s m^2}{4\pi}\l[-i\pi\coth\l(\frac{\pi C\bar\alpha_s}{2}\sqrt{\frac{m}{E}}\r) + \gamma_E\r]
\simeq -\frac{C\bar\alpha_s m^2}{4\pi}\l(\gamma_E - i\pi\r) \,,
\ee
which gives
\be
\l|G(\vv{0},E)\r|^2
\simeq \frac{C^2\bar\alpha_s^2 m^4}{16\pi^2}\l(\gamma_E^2 + \pi^2\r) \,.
\label{|G|2-thresh}
\ee
Using (\ref{sigma-annih-gen-sigmas}) with (\ref{|G|2-thresh}) and (\ref{sigma-ImG/Phi2}) with (\ref{ImG-thresh}), we obtain the branching ratio of the annihilation-like processes in the near-threshold continuum:
\be
\frac{\hat\sigma_{ab\to\alpha\beta\to AB}}{\hat\sigma_{ab\to\alpha\beta}}
\simeq \frac{\gamma_E^2 + \pi^2}{8\pi^3}\,C\bar\alpha_s\, \frac{D_A D_B}{(2J+1)D_{\alpha\beta}}\frac{\hat\sigma_0^{AB\to\alpha\beta}}{\Phi_2}m^2
\q \l[\;\times \frac{1}{2} \;\,{\rm if}\; A = B\;\r]\,.
\label{BR-cont}
\ee
With typical strong-interaction diagrams we will have $\hat\sigma_0^{AB\to\alpha\beta}/\Phi_2 \sim \alpha_s^2/m^2$ giving
\be
\frac{\hat\sigma_{ab\to\alpha\beta\to AB}}{\hat\sigma_{ab\to\alpha\beta}} \sim C\bar\alpha_s\alpha_s^2 \sim 10^{-3} \,.
\label{BR-cont-oom}
\ee

\subsection{The case of $t\bar t$}

Let us apply the results of this Appendix to computing the curves in Figure~\ref{fig-toponium}. The near-threshold tree-level $t\bar t$ production cross sections in the various spin and color channels are (see, e.g.,~\cite{Bonciani:1998vc})
\be
\l\{\hat\sigma_0^{gg\to \,^1S_0(\mathbf{1})}\,,\;
\hat\sigma_0^{gg\to \,^1S_0(\mathbf{8})}\,,\;
\hat\sigma_0^{q\bar q\to \,^3S_1(\mathbf{8})}\r\}
= \l\{\frac{1}{96}\,,\;\frac{5}{192}\,,\;\frac{1}{9}\r\}
\,\times\, \frac{\pi\alpha_s^2}{m_t^2}\beta \,.
\ee
From (\ref{sigma-ImG/Phi2}), the production cross sections for these three channels are
\be
\hat\sigma(E) = \frac{4\pi^2\alpha_s^2}{m_t^4} \times
\l\{\frac{1}{96}\,\mbox{Im}\,G^\mathbf{1}(\vv{0},E)\,,\;
\frac{5}{192}\,\mbox{Im}\,G^\mathbf{8}(\vv{0},E)\,,\;
\frac{1}{9}\,\mbox{Im}\,G^\mathbf{8}(\vv{0},E)\r\} \,,
\ee
and from (\ref{sigma-annih-gen-sigmas}) the annihilation cross sections (for $A,B = a,b$) are
\be
\hat\sigma^{\rm ann}(E) = \frac{\pi^3\alpha_s^4}{m_t^6} \times
\l\{\frac{1}{18}\,\l|G^\mathbf{1}(\vv{0},E)\r|^2\,,\;
\frac{25}{576}\,\l|G^\mathbf{8}(\vv{0},E)\r|^2\,,\;
\frac{2}{27}\,\l|G^\mathbf{8}(\vv{0},E)\r|^2\r\} \,.
\ee
In practice, when the potential is repulsive $\l|G(\vv{0},E)\r|^2$ is highly suppressed in the near-threshold region, so only the color-singlet contributes to the annihilation signal in the case of $t\bar t$ pairs.

For the near-threshold continuum, just for the color-singlet, (\ref{BR-cont}) gives the branching ratio
\be
\frac{\hat\sigma_{gg\to t\bar t\to gg}}{\hat\sigma_{gg\to t\bar t}}
\simeq \frac{\gamma_E^2 + \pi^2}{3\pi} C\bar\alpha_s\alpha_s^2
= 2.4 \times 10^{-3} \,.
\ee

\section{Resonance cross sections from decay rates}
Consider the process
\be
a\,b \to M_\Gamma \to A\,B\ldots \,,
\label{general-process}
\ee
in which partons $a$ and $b$ (quarks or gluons) produce a resonance of mass $M$, width $\Gamma$, spin $J$, in a specific color representation, that then decays into particles $A$ and $B$ (or more). The resonance propagator contributes the denominator
\be
\frac{1}{\l(\hat s-M^2\r)^2 + M^2\Gamma^2} \,.
\ee
The tensor structure of the propagator can be written as a sum of products of polarization vectors which will be absorbed in the decay rates that we will now discuss. The final part of the process (\ref{general-process}) is the same as in the decay $M \to A\,B$ (by rotational invariance, the total decay rate is the same for every polarization of $M$), so it contributes
\be
2M\,\Gamma_{M\to AB} \,.
\ee
Similarly, the initial part of the squared matrix element, summed (for the purpose of the derivation) over all possible directions for the incoming particles and over the polarizations of $a$, $b$ and $M$, is proportional to $\Gamma_{M\to ab}$. It is given by
\be
\frac{\l(2J+1\r)D_M}{2D_a\cdot 2D_b}\,\frac{4\pi}{\hat s}
\,2M\,\Gamma_{M\to ab}\,\qq\l[\;\times 2 \q{\rm if}\q a = b\;\r].
\ee
where we undid the integration over the different directions (which is trivial after summing over the polarizations), denoted
the dimension of the color representation of $X$ by $D_X$, and assumed $a$ and $b$ to have two helicities. This gives
\be
\hat\sigma_{ab\to M\to AB}(\hat s)
= \frac{4\pi\l(2J+1\r)D_M}{D_a D_b}{M^2\over\hat s}\frac{\Gamma_{M\to ab}\,\Gamma_{M\to AB}}{\l(\hat s-M^2\r)^2 + M^2\Gamma^2}\qq \l[\;\times 2 \q{\rm if}\q a = b\;\r] \,.
\ee
We should note, however, that the actual mass of the resonance is $\sqrt{\hat s}$ rather than $M$, and we need to make this replacement throughout the calculation:
\be
\hat\sigma_{ab\to M\to AB}(\hat s) = \frac{4\pi \l(2J+1\r) D_M}{D_a D_b}\frac{\Gamma_{M\to ab}\,\Gamma_{M\to AB}}{\l(\hat s-M^2\r)^2 + \hat s\,\Gamma^2}\qq \l[\;\times 2 \q{\rm if}\q a = b\;\r] \,,
\label{sigma-hat-general-partial}
\ee
where the widths $\Gamma_{M\to ab}$, $\Gamma_{M\to AB}$ and $\Gamma$ should be evaluated assuming resonance mass $\sqrt{\hat s}$ instead of $M$. In the narrow-width limit $\Gamma \ll M$, (\ref{sigma-hat-general-partial}) becomes
\be
\hat\sigma_{ab\to M\to AB}(\hat s) \simeq \frac{2\pi \l(2J+1\r) D_M}{D_a D_b}\frac{\Gamma_{M\to ab}\,\Gamma_{M\to AB}}{M\,\Gamma}\;2\pi\,\delta(\hat s - M^2)\qq \l[\;\times 2 \q{\rm if}\q a = b\;\r] \,,
\label{sigma-hat-general-partial-narrow}
\ee
and if we further sum over all the possible final states in (\ref{general-process}) this simplifies to
\be
\hat\sigma_{ab\to M}(\hat s) \simeq \frac{2\pi \l(2J+1\r) D_M}{D_a D_b}\, \frac{\Gamma_{M\to ab}}{M}\;2\pi\,\delta(\hat s - M^2)\qq \l[\;\times 2 \q{\rm if}\q a = b\;\r] \,.
\label{sigma-hat-general-total}
\ee
As expected, the production cross section is independent of the available decay modes.

\section{Algorithm for $t\bar t$ reconstruction\label{sec-ttbar-algorithm}}

This is the algorithm we use for the reconstruction of the $t\bar t$ pair in the semileptonic channel with an electron or a muon. It includes the most essential features of the algorithms studied by {\sc ATLAS}~\cite{ATL-PHYS-PUB-2006-033} and {\sc CMS}~\cite{CMS-PAS-TOP-09-009}. The efficiencies given below were obtained for signals of scalar and vector $600$ GeV resonances and the standard model background that was described in Section~\ref{sec-simulation}. The efficiencies are listed in the order scalar/vector/background.
\begin{enumerate}
\item We select events with exactly one electron or muon with $p_T > 35$ GeV and rapidity $|y| < 2.5$. Efficiency: $30\%/28\%/22\%$.
\item We save all jets with $p_T > 20$ GeV, but require to have at least 4 jets with $p_T > 30$ GeV. Efficiency: $75\%/57\%/64\%$.
\item We require that the 4 hardest jets have rapidities $|y| < 2.5$. Efficiency: $75\%/75\%/67\%$.
\item We start with reconstructing the leptonically decaying top. We compute the transverse momentum of the neutrino from the transverse momenta of the lepton and all the jets. By requiring $W$ to be on-shell, we obtain a quadratic equation for the longitudinal momentum of the neutrino, and keep both solutions. If the discriminant is negative, we try reducing neutrino's transverse momentum by up to $15\%$ till the solution is possible. Efficiency: $76\%/79\%/76\%$ (neutrino momentum reduction was used in $7\%/7\%/8\%$ of these cases).
\item We then find which of the jets best reproduces the mass of the top when combined with the $W$. We choose the neutrino momentum assignment that gives the better top mass accuracy. We require top mass accuracy of at least $10\%$. Efficiency: $87\%/87\%/88\%$.
\item To reconstruct the hadronically decaying top, we take the remaining jets and find the assignment of three of them that reconstructs the $W$ and the top most successfully, based on the criterion of minimizing the quantity
    \be
    e \equiv \sqrt{\l(\frac{\Delta m_W}{0.1 m_W}\r)^2 + \l(\frac{\Delta m_t}{0.2 m_t}\r)^2} \;,
    \ee
    where $\Delta m_W$ and $\Delta m_t$ are the differences between the reconstructed and the actual masses. We require that the best assignment has $e = e_{\rm best} < \sqrt{2}$ (efficiency: $47\%/41\%/46\%$) and the next-to-best assignment has $e > 3\,e_{\rm best}$ (efficiency: $24\%/34\%/26\%$).
\item To eliminate many of the events that were not reconstructed correctly, we require that the azimuthal angle $\phi$ between the two tops is at least $160^\circ$ (efficiency: $48\%/67\%/46\%$).
\end{enumerate}
The fraction of events that remains at the end of the algorithm is $0.6\%/0.8\%/0.34\%$ out of all the $t\bar t$ events.

\bibliography{ann}

\end{document}